\documentclass[]{aa}
\usepackage[varg]{txfonts}
\usepackage{xcolor}
\usepackage{twoopt}
\usepackage[%
  breaklinks = true,
  colorlinks = true,
  urlcolor   = blue,
  citecolor  = blue,
  linkcolor  = blue,
]{hyperref}

\bibpunct{(}{)}{;}{a}{}{,} 

\graphicspath{ {figures/} }

\makeatletter
 \newcommandtwoopt{\citeads}[3][][]{%
   \nonstopmode
   \href{http://adsabs.harvard.edu/abs/#3}%
        {\def\hyper@linkstart##1##2{}%
         \let\hyper@linkend\@empty\citealp[#1][#2]{#3}}
   \biblink{#3}{\href{http://adsabs.harvard.edu/abs/#3}{ADS}}%
   \errorstopmode}            
 \newcommandtwoopt{\citepads}[3][][]{%
   \nonstopmode
   \href{http://adsabs.harvard.edu/abs/#3}%
        {\def\hyper@linkstart##1##2{}%
         \let\hyper@linkend\@empty\citep[#1][#2]{#3}}
   \biblink{#3}{\href{http://adsabs.harvard.edu/abs/#3}{ADS}}
   \errorstopmode}            
 \newcommandtwoopt{\citetads}[3][][]{%
   \nonstopmode
   \href{http://adsabs.harvard.edu/abs/#3}
        {\def\hyper@linkstart##1##2{}%
         \let\hyper@linkend\@empty\citet[#1][#2]{#3}}
   \biblink{#3}{\href{http://adsabs.harvard.edu/abs/#3}{ADS}}%
   \errorstopmode}            
 \newcommandtwoopt{\citeyearads}[3][][]{%
   \nonstopmode
   \href{http://adsabs.harvard.edu/abs/#3}%
        {\def\hyper@linkstart##1##2{}%
         \let\hyper@linkend\@empty\citeyear[#1][#2]{#3}}
   \biblink{#3}{\href{http://adsabs.harvard.edu/abs/#3}{ADS}}%
   \errorstopmode}            
\makeatother
\makeatletter
\newcommand{\bibnote}[2]{\@namedef{#1note}{#2}}
\newcommand{\biblink}[2]{\@namedef{#1link}{#2}}
\makeatother

\newcommand{\CaIIHK}{\ion{Ca}{II}\,H\&K}
\newcommand{\CaII}{\ion{Ca}{II}}
\newcommand{\MgII}{\ion{Mg}{II}}
\newcommand{\KtwoV}{K$_{2\mathrm{V}}$}
\newcommand{\KtwoR}{K$_{2\mathrm{R}}$}
\newcommand{\Kone}{K$_1$}
\newcommand{\Ktwo}{K$_2$}
\newcommand{\Kthree}{K$_3$}

\begin{document}

\title{Three-dimensional modeling of the \CaIIHK\ lines in the solar atmosphere}

\author{
  Johan P. Bj{\o}rgen\inst{\ref{ISP}},
  Andrii V. Sukhorukov\inst{\ref{ISP},\ref{MAO}},
  Jorrit Leenaarts\inst{\ref{ISP}},
  Mats Carlsson\inst{\ref{UiO},\ref{RoCS}},
  Jaime de~la Cruz Rodr\'{\i}guez\inst{\ref{ISP}},
 G\"oran B. Scharmer \inst{\ref{ISP}}
  \and
   Viggo H. Hansteen\inst{\ref{UiO},\ref{RoCS}}
}

\institute{
  Institute for Solar Physics,
  Department of Astronomy, Stockholm University, AlbaNova University Centre,
  SE-106~91 Stockholm, Sweden
  \label{ISP}
  \and
 Main Astronomical Observatory, National Academy of Sciences of Ukraine,
  27 Akademika Zabolotnoho str., 03680 Kyiv, Ukraine
  \label{MAO}
  \and
   Institute of Theoretical Astrophysics, University of Oslo,
  PO Box 1029 Blindern, N-0315 Oslo, Norway
  \label{UiO}
  \and
Rosseland Centre for Solar Physics, University of Oslo, P.O. Box 1029 Blindern, N-0315 Oslo, Norway
\label{RoCS}
  }

\offprints{J. P. Bj{\o}rgen, \email{johan.bjorgen@astro.su.se}}

\date{Received; Accepted }

\abstract{
  CHROMIS, a new imaging spectrometer at the Swedish 1-m Solar Telescope (SST),
  can observe the chromosphere in the H and K lines of \ion{Ca}{II} at high
  spatial and spectral resolution.
  Accurate modeling as well as an understanding of the formation of these
  lines are needed to interpret the SST/CHROMIS observations.
  Such modeling is computationally challenging because these
  lines are influenced by strong departures from
  local thermodynamic equilibrium, three-dimensional
  radiative transfer, and partially coherent resonance scattering of photons.
}{
  We aim to model the \CaIIHK\ lines in 3D model atmospheres to understand their
  formation and to investigate their diagnostic potential for probing the
  chromosphere.
}{
  We model the synthetic spectrum of \CaII\ using the
  radiative transfer code Multi3D in three different
  radiation-magnetohydrodynamic model atmospheres computed with the Bifrost
  code.
  We classify synthetic intensity profiles according to their shapes and
  study how their features are related to the physical properties in the model
  atmospheres.
  We investigate whether the synthetic data reproduce the observed spatially-averaged line shapes,
   center-to-limb variation and compare with SST/CHROMIS images.
}{
 The spatially-averaged synthetic line profiles {show too low central emission peaks, and too small separation between the peaks}. The trends of the observed center-to-limb variation of the profiles properties are reproduced by the models. The \CaIIHK\ line profiles provide a temperature diagnostic of the temperature minimum and the temperature at the formation height of the emission peaks. The Doppler shift of the central depression is an excellent probe of the velocity in the upper chromosphere.
}
{}


\titlerunning{3D modeling of \ion{Ca}{II}~H\&K}
\authorrunning{Bj{\o}rgen et~al.}

\maketitle

\section{Introduction}

The resonance doublet of \ion{Ca}{II} represents the two strongest lines in the
visible solar spectrum, the H line at 3\,968.469~\AA\ and the K line at 3\,933.663~\AA\ (all wavelenghts are given in air for $\lambda>2000$~\AA).
Observations through the wings and the cores of these lines allow to investigate
the photosphere and the chromosphere.
The H and K lines of \ion{Ca}{II} share similar formation properties as the
h and k lines of \ion{Mg}{II}, typically showing wide damping wings, and central reversals in their cores.
As calcium is 18 times less abundant than magnesium in the solar atmosphere
%
\citepads{2009ARA&A..47..481A}, 
the H and K {line cores are formed lower in the chromosphere than the h and k cores}.

Most of the strongest and diagnostically-important chromospheric lines such as
the \ion{Mg}{II} h and k or \ion{H}{I} Ly-$\alpha$ and Ly-$\beta$ lines reside in the
ultraviolet part of the spectrum that is absorbed by the Earth's atmosphere, and therefore
they must be observed from space.
The \ion{Ca}{II} H and K lines are in the violet part of the visible spectrum
and can be observed with ground-based facilities such as the Swedish 1-m Solar
Telescope
\citepads[SST;]{2003SPIE.4853..341S}, 
the German Vacuum Tower Telescope
\citepads[e.g.,]{2007A&A...462..303T}, 
GREGOR
\citepads{2016A&A...596A...1S},
and the Dunn Solar Telescope
\citepads[e.g.,]{2009A&A...500.1239R}. 
%

The H and K line wings are formed in the photosphere with their opacity
following local thermodynamic equilibrium (LTE)
\citepads{2004A&A...413.1183R,  
          2012SoPh..280...83S}. 
They were used to obtain the temperature stratification of the upper photosphere
\citepads{2002A&A...389.1020R, 
          2009A&A...500.1239R, 
          2012A&A...548A.114H} 
and to investigate the reversed granulation both in observations and simulations
\citepads{2005A&A...431..687L}. 
%

The H and K line cores are formed in the chromosphere, and cover a narrow spectral range of $\sim0.4$~\AA.
So far, imaging observations in the cores have been performed with broad-band
filters having their transmission profiles 0.3--3~\AA\ wide
\citepads[e.g.,][]{1974SoPh...38...91Z,  
          2004A&A...413.1183R,  
          2007SoPh..243....3K,  
          2009A&A...500.1239R,  
          2009A&A...502..647P
          }.
Thus, previously-observed H and K core images were strongly contaminated with photospheric signal coming from the wings.

In August 2016 the new instrument CHROMIS was installed at the SST.
CHROMIS is an imaging spectrometer for the blue part of the spectrum designed as
a dual Fabry-P{\'e}rot filter system with a spectral transmission
profile of ${\sim}120$~m\AA\ width around 400~nm.
The system is optimized for a short integration time allowing to scan fast
through the line core with a high time cadence and minimal degradation caused by
the atmospheric turbulence.
By using image post-processing CHROMIS data can reach a diffraction-limited spatial resolution of $1.22 \lambda/\rm{D} \approx$ 0{\farcs}1 (or 73~km on the surface of the
Sun), which is close to the spatial resolution of today's magnetohydrodynamic
simulations of the chromosphere
\citepads{2016A&A...585A...4C}. 

CHROMIS allows to sample the inner wings and line-core of the H or K lines in $\sim10$~\,s with multiple frames and good signal-to-noise, in 2D images at the diffraction limit and at a high spectral resolution so that a clean chromospheric signal is obtained.

In the solar spectrum, the infrared triplet of \ion{Ca}{II} consists of three
strong lines at 8\,498.018~\AA, 8\,542.089~\AA, and 8\,662.140~\AA, whose cores
are formed in the chromosphere as well.
Among them, the \ion{Ca}{II} 8\,542~\AA\ line is the most studied and used {to investigate the magnetic field and temperature structure}
\citepads[see][and reference therein]
  {2008A&A...480..515C,  
   2013ApJ...764L..11D}. 

The 8\,542~\AA\ line is only weakly affected by horizontal radiative transfer
(3D) and partial redistribution (PRD) effects and can be modeled with a
modest computational effort {in one-dimensional (1D) models}
\citepads{2012A&A...543A..34D}. 

The H and K lines are formed much higher, in the
less dense upper chromosphere where 3D and PRD effects play an essential role in the line formation
%
\citepads{1953ZA.....31..282M}, 
than the infrared lines.
Previously, these lines have been modeled including effects of PRD, but only in
a one-dimensional (1D) radiative transfer approach
\citepads{1974SoPh...38..367V, 
          1975ApJ...199..724S, 
          1989A&A...213..360U, 
          1991A&A...250..220S, 
          2008A&A...484..503R}. 
Using a 3D non-LTE radiative transfer approach including
effects of PRD has become feasible recently with an upgrade of the Multi3D
code
\citepads{2017A&A...597A..46S}. 
Previously, the most accurate treatment of chromospheric lines was to model
features in the core and in the wings of lines separately using different
numerical codes like in
\citetads{2013ApJ...772...90L}
or
\citetads{2013ApJ...778..143P}.
The Multi3D code was used to model the line core including 3D radiative transfer
 but in the simplifying approximation of complete redistribution (CRD).
The RH code 
{\citepads{2001ApJ...557..389U}}
was used to model the wings of the line using a {1.5D} radiative transfer
approach but including PRD effects, which are essential in the inner wings.

In this paper, we use various model atmospheres (3D snapshots) computed with
the Bifrost code
\citepads{2011A&A...531A.154G}, 
to model the formation of the \ion{Ca}{II} H and K
lines in a full 3D non-LTE PRD approach using the Multi3D code.
We compare our calculated data with observations of a quiet Sun region taken by
SST/CHROMIS.

Section \ref{sec:observation} presents the observations taken with SST/CHROMIS.
We discuss the method and the setup of our computations as well as important
details of the PRD line transfer in Section~\ref{sec:modeling}.
We compare morphological properties of images as well as statistical
properties of the line profile features for our calculated and observed data
sets in Section~\ref{sec:comparison_obs}.
In Section \ref{sec:result_diagnostic}, we discuss
 how the observed features of the lines correlate with the properties of
the atmosphere and what observable diagnostics are the most useful to probe the
chromosphere.
In Section \ref{sec:discussion}, we conclude and suggest
how observations in the \ion{Ca}{II} H and K lines can be used to study the
chromosphere.

\section{Observations}
\label{sec:observation}

\begin{figure}
  \includegraphics[%
    width = \columnwidth,%
    trim  = 1pt 2pt 0pt 1pt,%
    clip  = true%
  ]{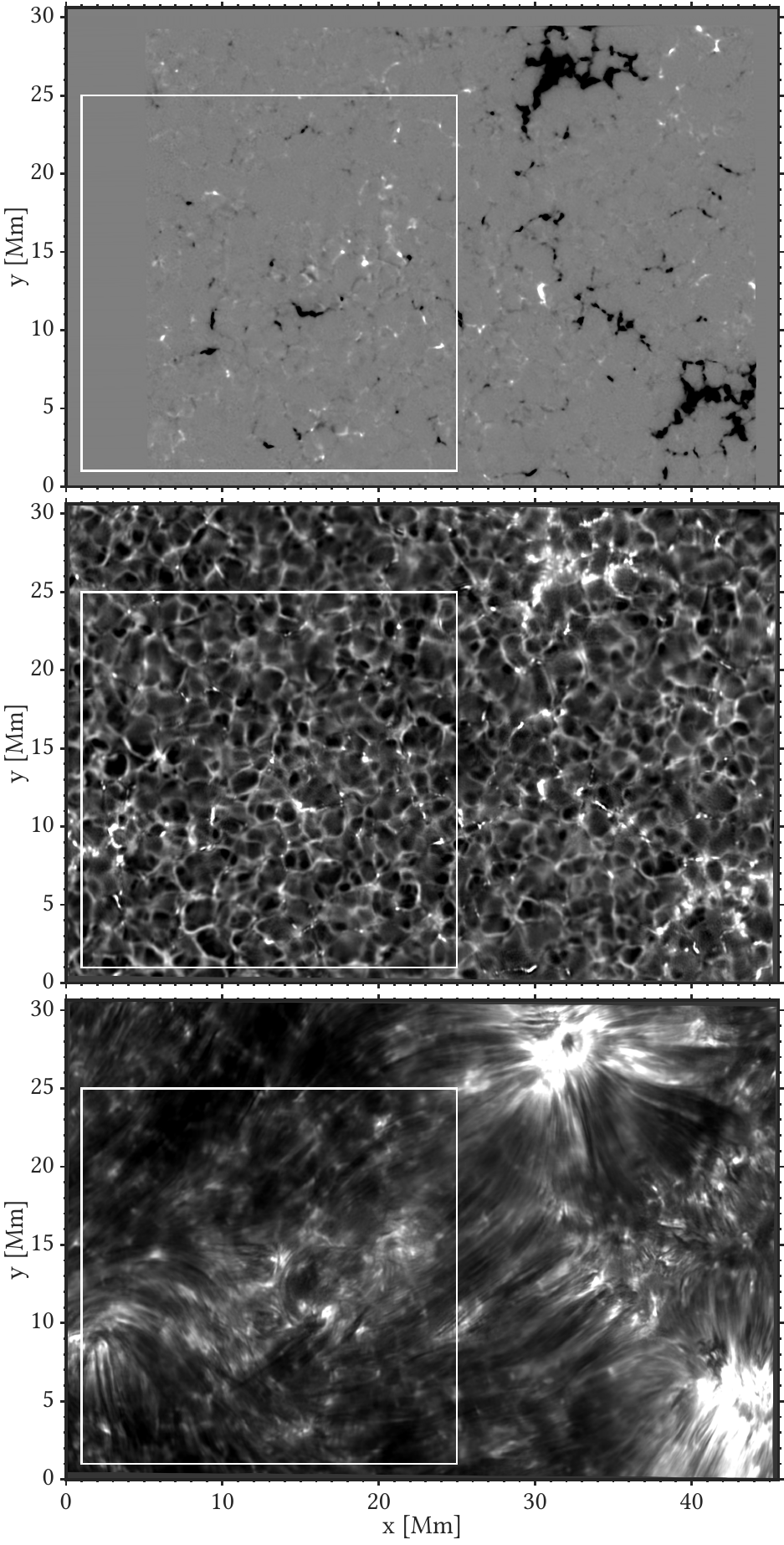}%
  \caption{Magnetogram (\emph{upper panel}) and \ion{Ca}{II}~K images (\emph{middle panel} and \emph{lower panel}) taken by SST/CHROMIS of the quiet Sun near the
    disk center ($\mu = 1.0$).
    \emph{Upper panel:}  vertical magnetic field obtained from Milne-Eddington inversion with Fe \textsc{i} taken by SST/CRISP. The color bar range is [$-300$~G, $300$~G].
    \emph{Middle panel:} red wing position at $\Delta\lambda = +1\,409$~m\AA.
    \emph{Lower panel:} line center position at $\Delta\lambda = 0$~m\AA.
    The white square outlines the region that we compare with our simulations
    (see Fig.\ \ref{fig:obs2}).}
  \label{fig:obs1}
\end{figure}

We use data observed with the CHROMIS instrument at the Swedish 1-m Solar
Telescope on October 12, 2016 at 10:46--10:56~UT.
The target was a quiet Sun region near the disk center at $\theta_x = 2$\arcsec,
$\theta_y = -38$\arcsec (Helioprojective-Cartesian coordinates).

The \ion{Ca}{II}~K line was sampled with 36 wavelength points covering
a 1.409~\AA\ interval around the line center at 3\,933.664~\AA\ with one
continuum point at 4\,000~\AA.
The line core was sampled within a $\pm 0.528$~\AA\ interval with 59~m\AA\ spacing.
The line wings were sampled outside the core interval up to $\pm 1.409$~\AA\ with
118~m\AA\ spacing. 
The camera was run at $80$ frames per second with an exposure time of $12$~ms;  a full line scan took 13 seconds. 
CHROMIS has a spectral transmission profile with a 120 m\AA\ FWHM, a field of view about 63'' $\times$ 42'' and a pixel size of 0\farcs0375.

The \ion{Ca}{II}~K data set is complemented with {observations of the magnetically sensitive} \ion{Fe}{I}
6\,302~\AA\ line taken simultaneously on the same target with the CRisp
Imaging SpectroPolarimeter
\citepads[CRISP;]{2008ApJ...689L..69S}. 
The \ion{Fe}{I} 6\,302~\AA\ was sampled with 16 wavelength points on a non-equidistant wavelength 
grid covering from $-1180$~m\AA\ to $+80$~m\AA\ around 6\,302~\AA. A full line scan took 8 
seconds. We acquired also simultaneous CRISP observations in \ion{Ca}{II}~8542 \AA\ and H-$\alpha$, so the total cadence is $37$ seconds. 

Final data sets were produced from the raw data using the CHROMISRED pipeline
(L{\"o}fdahl et al. in prep.) for the \ion{Ca}{II}~K observations and the CRISPRED pipeline
\citepads{2015A&A...573A..40D} 
for the \ion{Fe}{I} observations. The CHROMIS data where calibrated by scaling the spatially-averaged spectrum to an atlas profile.

Figure~\ref{fig:obs1} shows two images from the \ion{Ca}{II}~K
observations in the red wing and the core of the line.
The wing image shows the upper photosphere with a distinct
reversed granulation pattern and bright magnetic field concentrations
in intergranular lanes.
The core image shows the chromosphere covered with thin, elongated fibrils that appear
resolved at the spatial resolution of CHROMIS. In the upper panel we show the vertical component of the magnetic field vector, derived from a Milne-Eddington inversion
of the \ion{Fe}{i} photospheric data. These inversions were performed with a modified version of the 1D code presented in
\citetads{2015A&A...577A.140A}.
%

From the whole field of view we selected a square region of the quiet Sun,
outlined in Fig.~\ref{fig:obs1}.
This region matches the physical extent of our simulations and has a similar photospheric
magnetic field configuration.
We use observed data within this region for the comparison with our synthetic data.

\section{Modeling} \label{sec:modeling}

\subsection{Radiative transfer computations} \label{sec:radtrans}

We numerically solve the non-LTE radiative transfer problem with the latest
version of the Multi3D code
\citepads{2009ASPC..415...87L} 
in various model atmospheres discretized on a Cartesian three-dimensional (3D)
grid.

For a given model atom, the code simultaneously solves the system of statistical
equilibrium equations and integrates the radiative transfer equation at spectral
points covered by the bound-bound and bound-free transitions of the model atom.
The solution is computed by iteration until convergence using multilevel
accelerated $\Lambda$-iteration (M-ALI) with pre-conditioned radiative rates
following
\citetads{1991A&A...245..171R,  
          1992A&A...262..209R}. 
The method of short characteristics
\citepads{1987JQSRT..38..325O} %
is used to integrate the transfer equation.
Either linear of the 3rd-order hermitian
\citepads{2003ASPC..288....3A,2013A&A...549A.126I} 
interpolation is used to approximate the source function in the formal solution
of the transfer equation.
We use the 24-angle quadrature (set ``A4'') from
\citet{carlson1963}. 

The code allows to solve the radiative transfer equation either in
3D by taking into account the horizontal transfer or radiation, or in
the {1.5D} approximation by treating each vertical column as an independent plane-parallel
atmosphere.

By default, the code treats line scattering with complete redistribution (CRD).
We use a recent upgrade of the code
\citepads{2017A&A...597A..46S} 
that allows to treat resonance line scattering with partial redistribution (PRD)
as well as cross-redistribution (XRD).
For more details we refer to Section~\ref{sec:line-treatment}.

\subsection{Model atom}
\label{sec:model-atom}

\begin{figure}
  \includegraphics[width = \columnwidth]{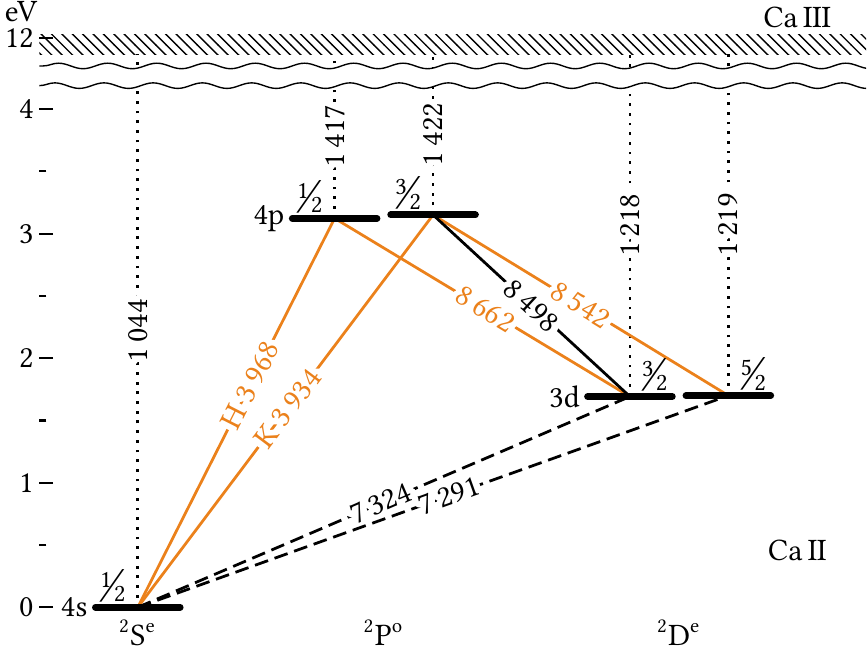}
  \caption{%
    Term diagram of the \ion{Ca}{II} model atom.
    Atomic levels (horizontal bars) are shown with their valence electron
    configuration $nl$ to the left, the total angular momentum $J$ on top (as a
    fraction), grouped by their term configuration $^{2S+1\!}L^P$ at the bottom
    row.
    Bound-bound permitted (solid lines) and forbidden (dashed lines) transitions
    connect the levels.
    PRD transitions are orange.
    Bound-free transitions (dotted lines) connect their levels to the
    \ion{Ca}{II} continuum (hashed area).
    For all transitions, line center or threshold wavelengths are given in
    {\AA}ngstr\"oms.%
  }
  \label{fig:model-atom}
\end{figure}

We used a five-level plus continuum model atom of the \ion{Ca}{II} ion
illustrated in Fig.\ \ref{fig:model-atom}.
It contains the lowest levels of \ion{Ca}{II} that are sufficient to represent
the physics of formation for the H, K, and T lines together.
The properties of the atomic levels are from the NIST Atomic Spectra Database
following
\citetads{1985aeli.book.....S} 
for \ion{Ca}{II} and
\citetads{1956ArF...10....553} 
for \ion{Ca}{III}.

Transition probabilities for the permitted transitions (H, K, and T) are from
\citetads{1989PhRvA..39.4880T}. 
To ensure the correct population of the 3d~$^2$D$^\text{e}$ term, we added the
3d~$^2$D$^\text{e}$ -- 4s~$^2$S$^\text{e}$ multiplet with two forbidden lines at
7\,291.4714~\AA\ and 7\,323.8901~\AA\ with transition probabilities from
\citetads{1951ApJ...114..469O}. 
Both forbidden lines are present, although blended, in the solar spectrum
\citepads{1968CR....266..110G,  
          1969SoPh....7...11L,  
          1969SoPh...10..311L,  
          1975SoPh...43....9S,  
          1974SoPh...36...25D}. 
The broadening parameters of all lines are from the Vienna Atomic Line Database
\citepads{1995A&AS..112..525P, 
          1999A&AS..138..119K} 
among which the van der Waals parameters are taken from
\citetads{2000A&AS..142..467B}. 

Photoionization cross-sections for the bound-free transitions are from the
TOPBase server of the Opacity Project
\citepads{1994MNRAS.266..805S}.
The original cross-sections are sampled on a very fine $10^3$--$10^4$-point grid
of frequencies with well-resolved resonance and autoionization transitions.
For each atomic level in the model, we smoothed and downsampled the original
data to ${\sim}$30-point grid following
\citetads{1998ApJS..118..259B} 
and
\citetads{2003ApJS..147..363A}. 

Bound-bound {electron} collisional rates are composed of data from
\citetads{2007A&A...469.1203M} 
and extrapolations following
\citetads{1992A&A...254..436B}. 
Bound-free collisional rates are either from
\citetads{1985A&AS...60..425A} 
for the ground 4s~$^2$S$^\text{e}$ level or from the general formula provided by
\citetads{1983MNRAS.203.1269B} 
for the excited levels.
We also include collisional autoionization
\citepads{1985A&AS...60..425A} 
and dielectronic recombination
\citepads{1982ApJS...48...95S}. 

We adopt a standard atomic weight of the Ca atom, 40.078~{a.m.u.}\ for a
 mixture of 96.9\% $^{40}$Ca, 2.1\% $^{44}$Ca, and 1.0\%
$^{42,43,46,48}$Ca isotopes.
The solar abundance of Ca is taken to be~6.34 on the standard $[\mathrm{H}]=12.00$ scale
\citep{2009ARA&A..47..481A}. 
%

\subsection{Model atmospheres}
\label{sec:model_atmosphere}

\begin{figure}
  \includegraphics[width = \columnwidth]{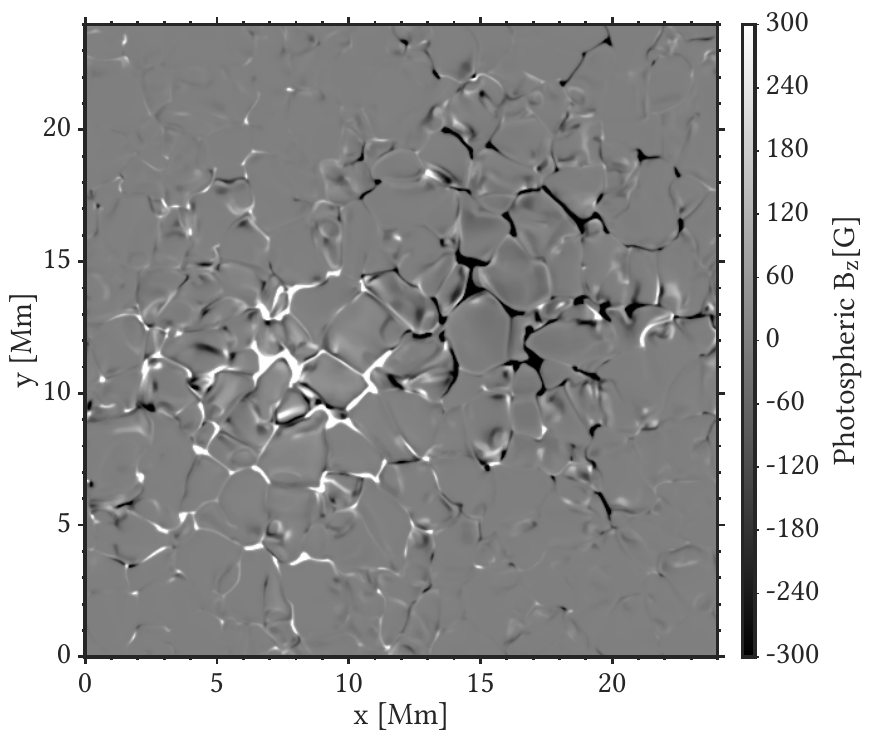}
  \caption{ {The vertical magnetic field strength in the photosphere in Model 2, at the height where the average optical depth at 5000 \AA\ is unity.}
  }
  \label{fig:model-atomsphere-magnetic}
\end{figure}

As model atmospheres we use three snapshots from three
different radiation-magnetohydrodynamic (R-MHD) numerical simulations done with
the Bifrost code
\citepads{2011A&A...531A.154G}. 
All three runs simulated a bipolar magnetic region, {which consists of two magnetic polarity patches separated by 8 Mm (illustrated for Model 2 in Fig.~\ref{fig:model-atomsphere-magnetic}). The region is} similar to an enhanced
network with an unsigned magnetic field strength of 50~G in the photosphere.
In all three cases, the simulation box has the same physical size of
24~Mm~$\times$~24~Mm~$\times$~16.9~Mm spanning from the top of the convection
zone up to the corona.
The models differ in the spatial resolutions of their coordinate grids and in
the equations of state (EoS) used for the initial R-MHD setup.
We refer to these models as Model~1, Model~2, and Model~3.
We selected and prepared all three models so that, compared to the
diffraction-limited spatial resolution of SST/CHROMIS, their horizontal grid
spacing is larger for Model~1, and is smaller for Model~2 and Model~3.

%
%
Model~1 is based on the public Bifrost model atmosphere published by
\citetads{2016A&A...585A...4C}. 
We took a snapshot at t=3\,850~s of solar time.
{The simulation used an EoS that includes the effects of non-equilibrium ionization of hydrogen}
{\citepads{2007A&A...473..625L}}. 
The original model has $504 \times 504 \times 496$ grid points within the full
physical extent of the simulations.
We reduced the grid size of this model to save computational time.
We clipped the vertical range of heights to $-$0.48\ldots$+$14.2~Mm keeping only
formation heights of the \ion{Ca}{II} spectrum.
We also halved the horizontal grid resolution by removing every other point in
the XY-direction.
The final model has $252 \times 252 \times 440$ grid points with a uniform
horizontal grid spacing of 95~km and a vertical grid spacing ranging from 19~km in
the photosphere and the chromosphere to 96~km in
the corona.

%
%
Model~2 was made using the same initial setup as Model~1.
There are two differences.
First, this simulation was done using a different EoS that includes effects of
the non-equilibrium ionization of hydrogen
and helium
\citepads{2014ApJ...784...30G,2016ApJ...817..125G}. 
Second, we took this snapshot at a different moment of simulation time, 780~s
after the running code was switched from the LTE~EoS to the non-equilibrium EoS
of hydrogen and helium.
We clipped the vertical range of heights to $-0.53$\ldots$+$6.6~Mm, but kept the
original horizontal grid resolution.
The final model has $504 \times 504 \times 336$ grid points with a uniform
horizontal grid spacing of 48~km.

%
%
Model~3 has a different EoS setup.
The ionization and the recombination of hydrogen and helium were treated in
instantaneous LTE{, which means that the atomic number densities follow the Saha-Boltzmann equations.}
The original size of this model grid is $768 \times 768 \times 768$ points.
We clipped the vertical range of heights to the same range as in Model~2 and we
kept the original horizontal resolution.
The final model has $768 \times 768 \times 476$ grid points with the uniform
horizontal grid spacing of 31~km and a vertical grid spacing from 13~km in
the photosphere and the chromosphere to 27~km in
the corona.

For a comparison of the different EoS effect on the temperature stratification, 
we refer the reader to 
\citetads{2016ApJ...817..125G}.  
%

\subsection{Line treatment in CRD, PRD, and XRD}
\label{sec:line-treatment}

Contrary to many photospheric lines, which can be modeled assuming photon
scattering with complete redistribution (CRD), the resonance doublet as well as
the infrared triplet of \ion{Ca}{II} are formed in the chromosphere and require
a more accurate treatment of resonance photon scattering with partial
redistribution. {In PRD, the frequency and direction of the ingoing and outgoing photon in a scattering event can be correlated. To the contrary, in CRD they are independent.}

In addition, as all the lines share the same upper term 4p~$^2$P$^\text{o}$ and
have either sharp (4s~$^2$S$^\text{e}$ for H and K) or metastable
(3d~$^2$D$^\text{e}$ for infrared triplet) lower terms, they all are affected by resonance
Raman scattering of photons, often called ``cross-redistribution'' (XRD).
Thus, a photon absorbed in one of the H, K, or infrared triplet lines can be emitted in the
same line (resonance scattering) or in one of the other lines (resonance Raman
scattering).
A classical example of cross-redistribution in astrophysics is the formation of
the \ion{H}{I}~Ly-$\beta$ line
\citepads{1995ApJ...455..376H}, 
which is interlocked with the H-$\alpha$ line.

Following
\citetads{1989A&A...213..360U}, 
we tested the formation of all the five lines either in CRD or PRD, with or
without XRD using various 1D models of the solar atmosphere.
We found that PRD is essential for the H and K lines, but less
important for the 8\,542~\AA\ and 8\,662~\AA\ lines.
Cross-redistribution has very weak effect on the intensity profiles of the infrared triplet
lines, but generally makes 2\%\,--\,10\% intensity difference in the inner
wings of the H and K lines.
The 8\,498~\AA\ line has the smallest transition probability and is formed
mostly in the photosphere.
It shows very little effects of PRD and makes no contribution with XRD to its
subordinates, the K and the 8\,542~\AA\ lines.
We treat the 8\,498~\AA\ line in CRD reducing the total computational time by
10\%. {Including XRD increases the total computational time by 35 \% compared to PRD.}
We treat the H, K, 8\,542~\AA, and 8\,662~\AA\ lines with XRD (orange in
Fig.~\ref{fig:model-atom}).
Thus, photons can be either scattered resonantly in each of these lines, being
absorbed and emitted in the same transition, or cross-redistributed in the
{3\,968}~\AA${}\leftrightarrow{}$8\,662~\AA\ as well as
{3\,934}~\AA${}\leftrightarrow{}$8\,542~\AA\ cascades, being absorbed in one and
emitted in the other transition.

\begin{figure}
  \includegraphics[%
    width = \columnwidth,%
    trim  = 1pt 2pt 1pt 2pt,%
    clip  = true%
  ]{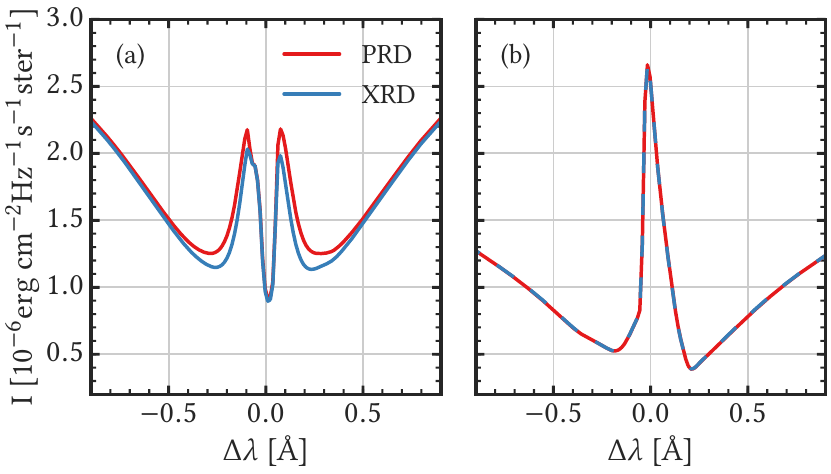}%
  \caption{Example profiles showing an effect (\emph{left}) and no effect
    (\emph{right}) of cross-redistribution (XRD, blue line) on the
    \ion{Ca}{II}~K intensity computed in 3D, extracted from two columns from Model~2 {at $\mu=1$ }.
    The same profiles computed with only partial redistribution but not cross-redistribution (PRD,
    red line) are given for comparison.}
  \label{fig:figure_xrd}
\end{figure}

Cross-redistribution provides an extra escape route for photons absorbed
in the H and K lines at heights where the subordinate infrared triplet lines are formed.
Thus, mostly the inner wings and the outer slopes of the peaks of the H and K
lines are affected, but not the cores as at those heights the infrared triplet lines are
optically thin and scattering redistribution is dominated by thermal motions
in the line cores.
Figure~\ref{fig:figure_xrd} illustrates this effect in the synthetic profiles of
the K line.
Normally, XRD slightly decreases the intensity in the inner wings without much
center-to-limb variation (see Fig.\ \ref{fig:center_2_limb}).



\subsection{Effects of {1.5D}/3D~RT and CRD/PRD/XRD}

\begin{figure*}
  \begin{minipage}{\textwidth}
    \includegraphics[%
      width = \textwidth,%
      trim  = 1pt 0pt 1pt 6pt,%
      clip  = true%
    ]{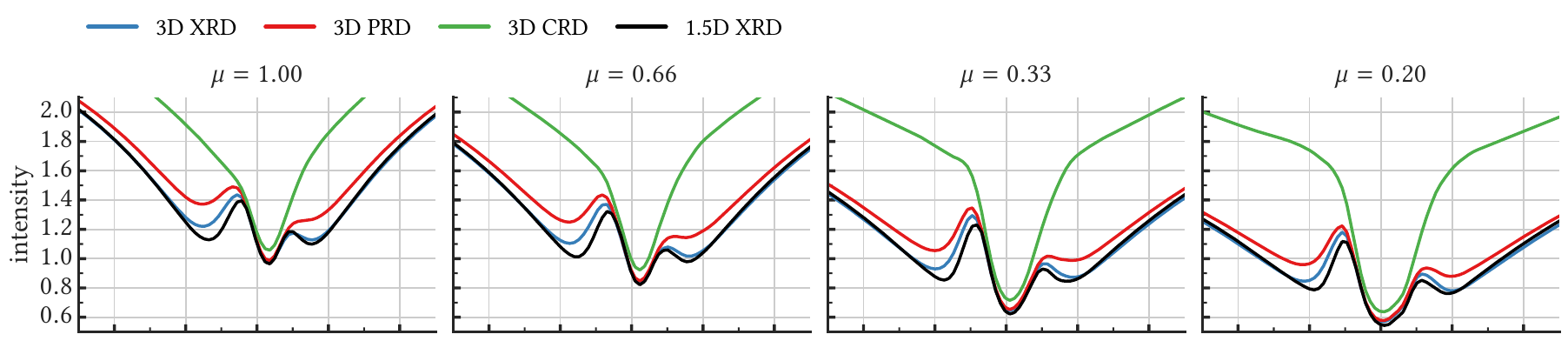}%
    \\
    \includegraphics[%
      width = \textwidth,%
      trim  = 1pt 3pt 1pt 4pt,%
      clip  = true%
    ]{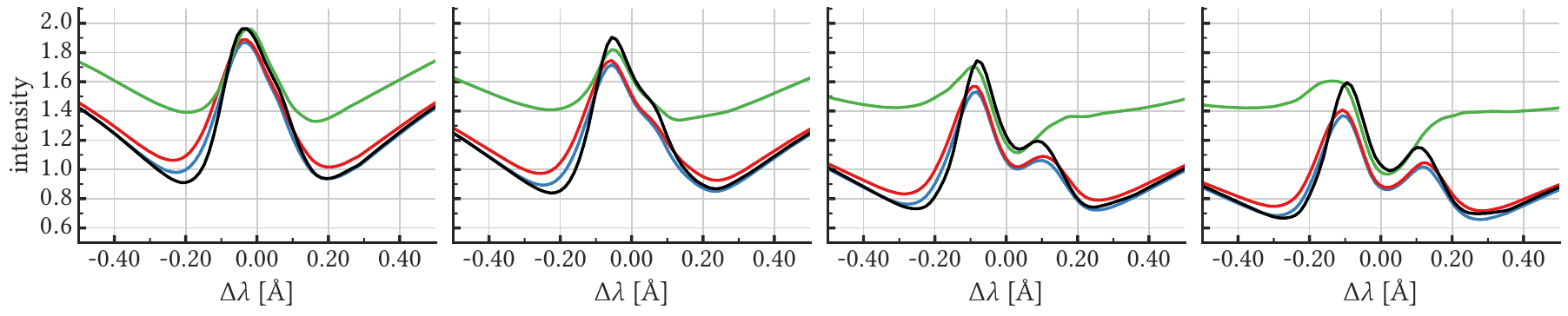}%
  \end{minipage}%
 \caption{%
    Spatially-averaged intensity profiles of the \ion{Ca}{II}~K line computed
    using Model~1 (\emph{upper row}) or Model~2 (\textit{lower row}) at four
    different angles $\mu = 1.0$, 0.66, 0.33, and 0.2 (columns from left to right).
    Results are given for 3D~XRD (blue), 3D~PRD (red), 3D~CRD (green), and
    {1.5D}~XRD (black).
    The intensity units are $10^{-6}$ erg s$^{-1}$ cm$^{-2}$ Hz$^{-1}$ ster$^{-1}$.%
  }
  \label{fig:center_2_limb}
\end{figure*}

\begin{figure*}
  \includegraphics[%
    width = \textwidth,%
    trim = 1pt 2pt 1pt 2pt,%
    clip = true%
  ]{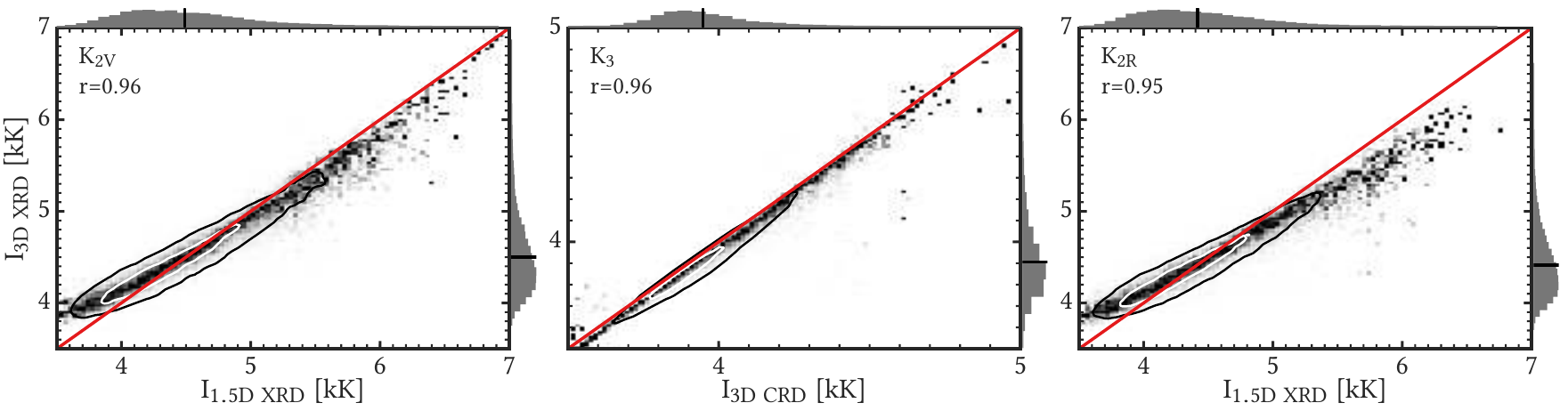}
  \caption{
    Joint-PDF distributions of the brightness temperature in K$_\text{2V}$
    (\emph{left}), K$_3$ (\emph{center}), and K$_\text{2R}$ (\emph{right}) for
    different combinations of the radiative transfer geometry ({1.5D} or 3D) and the
    line scattering (CRD or XRD).
    Intensities are calculated using Model~1 atmosphere at $\mu = 1$.
    The red line is $y = x$.
    The correlation coefficient $r$ is given on each panel.
    Contours encompass  50\% (white) and 90\% (black) of the pixels.
    Each column in the Joint PDF is scaled to maximum contrast.
    The short black lines at the univariate histograms
    (\emph{each panel's top and right}) show mean values.%
  }
  \label{fig:fig1_1dprd_vs_3dprd}
\end{figure*}

Because they could not perform non-LTE 3D PRD computations,
\citetads{2013ApJ...772...89L} 
modeled the \MgII~h\&k lines  with different treatments
in the core and in the wing parts of the profile:

In the cores of resonance lines, the redistribution is close to CRD as it is
controlled by random frequency shifts owing to thermal (Doppler) motions, which destroys the frequency-coherency of the scattering.
As resonance lines are strong and highly scattering, their cores are formed higher
up in the chromosphere where the effects of horizontal radiative transfer become
essential, and three-dimensional radiative transfer must be applied.
Therefore, a 3D~CRD treatment is reasonably accurate for the cores of such lines.

In the wings of resonance lines, PRD effects are more important because of the
radiative damping is much larger than collisional damping.
Because the line wings are formed relatively deep in the atmosphere and the effect of horizontal radiative transfer is small there, a {1.5D}~PRD treatment can be used to approximate the wing intensity for such lines.

\citetads{2017A&A...597A..46S} presented a method to perform radiative transfer computations in 3D non-LTE including PRD. Therefore we test the influence on the \CaIIHK\ lines of the simplifying assumptions of {1.5D}~XRD, 3D~CRD, and 3D~PRD compared to the most
accurate treatment of 3D~XRD.

\citetads{1989A&A...213..360U} 
tested whether the inclusion of cross-redistribution influences the intensity and center-to-limb variation of the \CaII~H line in a 1D model of the solar atmosphere. Here we perform a similar comparison, but now for a 3D atmosphere and including 3D radiative transfer.
Figure~\ref{fig:center_2_limb} shows the center-to-limb variation of
spatially-averaged intensity profiles of the K line treated accurately in 3D~XRD
and approximately in {1.5D}~XRD, 3D~CRD, and 3D~PRD.
We note that  3D effects are dominant in the line core although some
coherency is still present as there is a small intensity difference compared to
3D~XRD.
Outside of the line core, the redistribution effects become dominant and
3D~CRD produces large errors in the wings. The 3D~CRD reach the redistribution intensities at $\Delta \lambda \approx \pm 3 \AA$.
The redistribution effects increase towards the limb as the difference between
the 3D~CRD and 3D~XRD is increasing.
The {1.5D}~RT approximation is accurate in the outer wings (at more than $\sim 0.3$~\AA\ from line center)  of the line but is not at
the K$_1$ minima and K$_2$ peaks {(See section \ref{sec:line_features} for a definition of K$_1$ and K$_2$)}.
On average, the cross-redistribution decreases the intensity in the inner line wings
by 5--10\%.
%

%

Figure~\ref{fig:fig1_1dprd_vs_3dprd} shows how the 3D~XRD intensities
are related to the approximate {1.5D}~XRD intensities at the emission peaks (K$_2$, see Section~\ref{sec:line_features})  and the
approximate 3D~CRD intensities at the central line depression (K$_3$) for all profiles in Model~1.
At the core, the accurate and the approximate intensities are linearly related
and the 3D~CRD approximation overestimates radiation temperatures by less than
50~K.
At the emission peaks, there is a saturation effect in {1.5D}~XRD depending on the
range of observed intensities.
Below 4.5~kK intensities are underestimated by 300~K, above 5~kK intensities are
overestimated by 200~K, in between intensities are accurate.
This is similar to what was observed for the \ion{Mg}{II} h and k lines
\citepads{2017A&A...597A..46S}, 
although these lines are not affected by cross-redistribution.

Therefore, the K line intensity can be accurately modeled only if the effects of
3D radiative transfer and XRD are considered together.
The same is true for the H line.

Throughout the rest of this paper, we derive our results from the synthetic \CaIIHK\ data that were computed with the combined effects of non-LTE, 3D radiative transfer
with XRD.

\subsection{Line profile features} \label{sec:line_features}

The diagnostic properties of the \ion{Ca}{II} H and K lines can be investigated
with techniques of various complexity.
The easiest technique is to search how a single line parameter, for example
an equivalent width or a central depth, is related to some general property of
the model atmosphere, which is computationally and practically easy but is only of limited use.
The most elaborate technique would be solving the inverse problem of radiative
transfer in the H and K lines to restore the whole structure of the atmosphere
at formation heights of the lines. This is in principle possible in the {1.5D} approach
\citepads{2016ApJ...830L..30D},
where each pixel is treated as an independent atmosphere, but not with full 3D radiative transfer.

We choose a procedure of intermediate complexity similar to what was done by
\citetads{2013ApJ...772...90L}, 
as the \ion{Mg}{II} h and k and the \ion{Ca}{II} H and K lines are formed in a
similar way and have similar line profile shapes.
First, we synthesize the H and K intensity profiles, next we
classify the synthetic profile features, and finally we correlate the properties of the
features with the parameters of the model atmosphere at their formation heights.

\begin{figure}
  \includegraphics[%
    width = \columnwidth,%
    trim = 2pt 2pt 2pt 2pt,%
    clip = true%
  ]{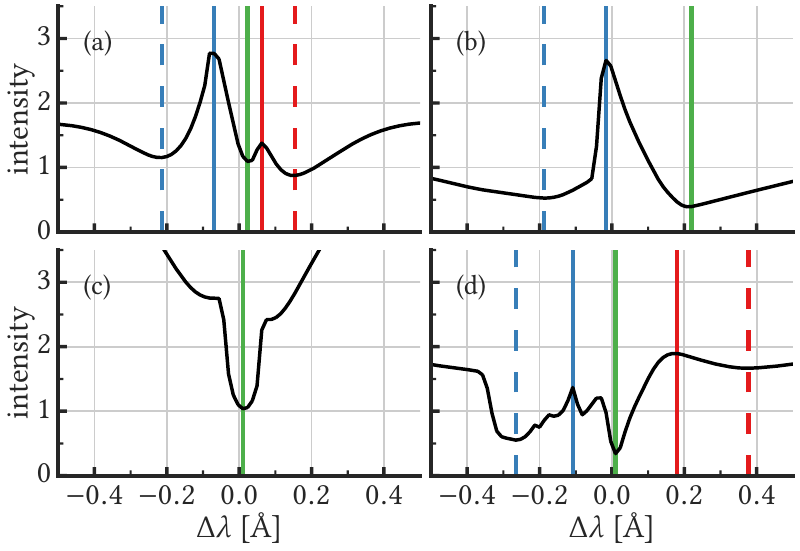}%
  \caption{Feature classification examples for the \ion{Ca}{II}~K line computed
    using Model~2.
    The intensity profile (solid black) is crossed by vertical lines marking
    K$_\mathrm{1V}$ (dashed blue), K$_\mathrm{2V}$ (solid blue), K$_3$ (solid
    green), K$_\mathrm{2R}$ (solid red), K$_\mathrm{1R}$ (dashed red) features.
    Panel a): standard profile with two emission peaks.
    Panel b): single-peaked emission.
    Panel c): pure absorption.
    Panel d): complex profile with multiple emission and absorption features.
    Intensity units are $10^{-6}$ erg cm$^{-2}$ Hz$^{-1}$ s$^{-1}$ ster$^{-1}$.}
  \label{fig:figure_1}
\end{figure}

We classify the profile features for the H and K lines using the notation system
introduced by
\citetads{1904ApJ....19...41H} 
for ''standard'' \CaII~K line-core profiles with two emission peaks close to the line core. Arriving from the short-wavelength side the first minimum is denoted as  K$_\mathrm{1V}$, the first emission peak is K$_\mathrm{2V}$, the central minimum is K$_3$, the second peak K$_\mathrm{2R}$ and the third minimum is K$_\mathrm{1R}$. The H line is characterized similarly.

We wrote a feature-finding code to automatically classify the line profiles in each pixel of each of the three models. Standard profiles (see Fig.~\ref{fig:figure_1}a), follow the classification described above.

If we find only one emission peak (as in Fig.~\ref{fig:figure_1}b), we assign
it either to the 2V- or the 2R-feature depending on which side it is with
respect to the nominal line center. We find that quite often the 3-feature can be mistaken for a 1-feature if the true line core (i.e., the wavelength with the largest $\tau_\nu=1$ height) is hidden
in the slope of the single 2-feature due to a strong velocity gradient. We therefore assign the 3-feature to the lowest intensity minimum next to the emission peak.
The other remaining minimum is the 1-feature. If the total shift of both the minima off the line center is more than
75~km\,s$^{-1}$, then there is no 3-feature and the minima are the 1V- and
1R-features. In 1\% --2\% of the single-peaked profiles the wavelength of the peak actually has the largest $\tau_\nu=1$ height. This happens when the source function is monotonically increasing with height in the chromosphere. This is rare in our quiet-Sun like atmosphere models, but might be more common in simulations with stronger magnetic activity.

If the profile has only one minimum and no peaks as in  Fig.~\ref{fig:figure_1}c then we only assign the 3-feature.

The most common complication is when the profile has more than two emission
peaks as in Fig.~\ref{fig:figure_1}d. We then assign the features based on rules that were empirically determined to give a reasonable result.

We note that the feature-finding algorithm still produces many incorrect identifications. Averaged over all three model atmospheres, we got the following fractions of the
\ion{Ca}{II} profile types: 1\% pure absorption, 16.5\% one emission peak,
61\% two emission peaks, 21.5\% several emission peaks.
Model~2 has the strongest velocity gradients in the chromosphere and produces
80\% more many-peaked profiles than Model~1 or Model~3 do.
As the classifying algorithm is ambiguous for many-peaked profiles, we have the
largest uncertainties with Model~2.

\subsection{Synthetic data degradation} \label{sec:datadegradation}

To compare our observations with our computations, we degraded the synthetic
\ion{Ca}{II}~K data to match the spatial and spectral resolution as
well as the wavelength sampling of SST/CHROMIS.

First, we convolved each synthetic image with the spatial PSF of the instrument,
a 2D Gaussian kernel having $\text{FWHM} = 0\farcs1$.
Second, we convolved each synthetic spectral profile with the measured
transmission profile of the CHROMIS filter having $\text{FWHM} = 120$~m\AA\ at
3\,930~\AA.
Third, we binned synthetic images to match the spatial pixel size of the
instrumental CCD chip, 0\farcs0375.
The physical extent of the synthetic image, 24~Mm${}\times{}$24~Mm, maps to
$884 \times 884$ pixels on the CCD chip.
Fourth, we sampled synthetic spectral profiles at wavelengths corresponding to
the wavelength grid without the continuum point along one scan of the CHROMIS
observations (see Sect.\ \ref{sec:observation}).
For each of the three atmosphere models. the resulting data is an array of $884 \times 884 \times 35$ values
along the $(\text{X}, \text{Y}, \lambda)$-directions.

We do not include any degradation of the synthetic data to for instrumental straylight or residual effects of
atmospheric turbulence. We therefore expect the synthetic images to have a considerably higher contrast than
the observations.

\section{Formation of \CaIIHK}
\label{sec:caiiHK}

{In this section we will discuss how the \CaIIHK\  are formed and display four exemplary line profiles from our 3D XRD computations. We only discuss the K line, since the  H line forms in the same way.}

\subsection{An illustration of the formation of \CaIIHK} \label{sec:formation_emission_peaks}

\begin{figure*}
  \sidecaption
  \begin{minipage}[b]{120mm}
  \includegraphics[width = \columnwidth]{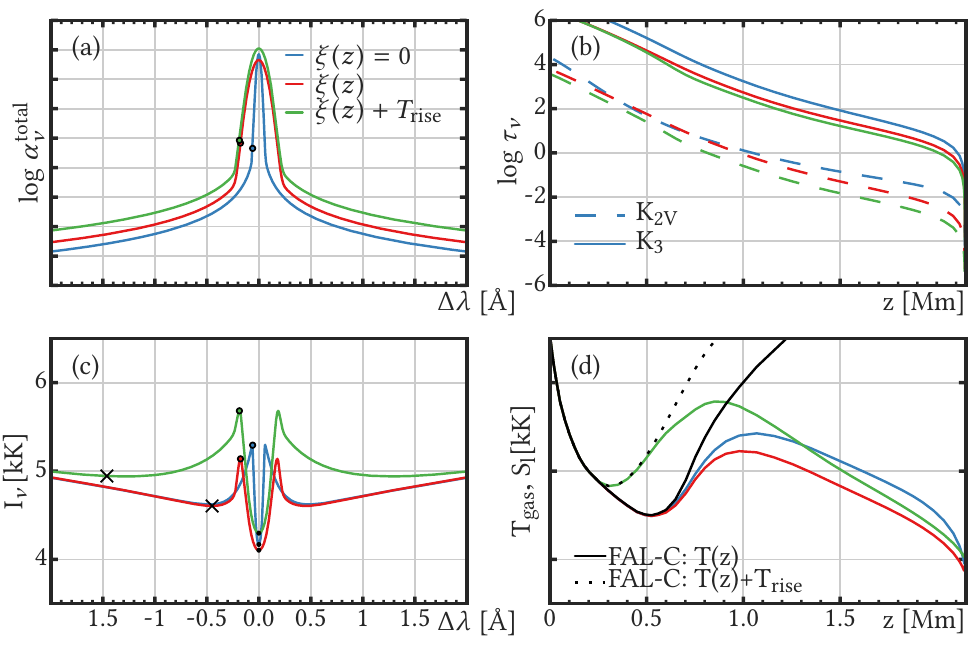}
  \end{minipage}%
  \caption{%
{A four-panel Eddington-Barbier line formation diagram for the \CaII\ K from the FAL-C model atmosphere in CRD. The computations are performed with three different cases: without microturbulence ($\xi(z) = 0~\rm{km}\,\rm{s}^{-1}$) (blue line), microturbulence ($\xi(z) = 5~\rm{km}\,\rm{s}^{-1}$) (red line), and with a deeper temperature rise and microturbulence (green line).
Panel (a) shows the total opacity as function of wavelength at the formation height of the peaks; panel (b) shows the optical depth as function of height at the wavelength of at K$_{\rm{2V}}$ (dashed) and K$_3$\ (solid); panel (c) shows the vertically emergent intensity for the K line, and panel (d) shows the gas temperature as a function of height (solid and dashed black) with the line source function (solid colored).  The cross marks the position of K$_{1\rm{V}}$ in panel (c) and the dot marks the position of the K$_{2\rm{V}}$ in panel (a) and K$_{2\rm{V}}$/K$_{3}$  in panel (c).}}
\label{fig:eb_four_panel}
\end{figure*}

{
The formation of the \CaIIHK\ lines are severely complicated by velocity fields, PRD/XRD effects, and the highly in-homogeneous temperature structure of the solar chromosphere (examples are shown in Section \ref{sec:intensity-formation}). To set the stage, we use the 1D FAL-C model atmosphere
\citepads{1993ApJ...406..319F},
to illustrate the basic formation of the \CaII\ K line in a four-panel Eddington-Barbier diagram.
}
{
 PRD adds an extra complexity to the analysis, by making the line source function frequency-dependent. To keep things somewhat simpler, we use the CRD approximation here, which means that the line source function is independent of frequency.}
 
 {Figure \ref{fig:eb_four_panel} shows $2\times2$ formation diagrams following
 \citet{2003rtsa.book.....R}
 for three different computations: the FAL-C temperature structure without microturbulence and with a constant microturbulence (5 km\,s$^{-1}$), and a computation with constant microturbulence (5 km\,s$^{-1}$), but with a modified temperature structure. 
The microturbulence, $\xi$, is an ad-hoc parameter used to broaden the spectral lines to fit the observed ones. We note that the model atmospheres from Bifrost do not include microturbulence.
}

{
Panel (c) in Fig. \ref{fig:eb_four_panel} shows the vertical emergent intensity for the \CaII\ K line. Panel (b) shows the optical depth as function of height at the wavelengths of the K$_{2\rm{V}}$ and  K$_{3}$ features.  The emission peak, K$_{2\rm{V}}$, is formed at the maximum of the line source function, which is shown in panel (d). The line source function is only partially coupled to the chromospheric temperature rise. The K$_{3}$ feature is formed at the largest formation height at ~1.9 Mm for all the cases. 
}
{
Panel (a) shows the broadening effect on the extinction profile from the microturbulence. Panel (c) shows that the microturbulence (5 km s$^{-1}$ in this case) increases the K$_{2}$ separation by a factor $3$ compared to the case without microturbulence.}
{
Panel (d) shows two different temperature stratifications, illustrating how the depth where the chromospheric temperature rise is located affects the emergent line profile. The K$_2$ separation increases slightly and the K$_1$ location shifts outward to $\Delta \lambda = \pm 1.5$ \AA\ with the deeper temperature rise shown in panel (c).
}
%

\subsection{Analysis of line intensity formation}
\label{sec:intensity-formation}


\begin{figure*}[htb!]
  \begin{minipage}{\textwidth}
    \includegraphics[%
      width = 0.5\textwidth,%
      trim = 2pt 2pt 2pt 3pt,%
      clip = true%
    ]{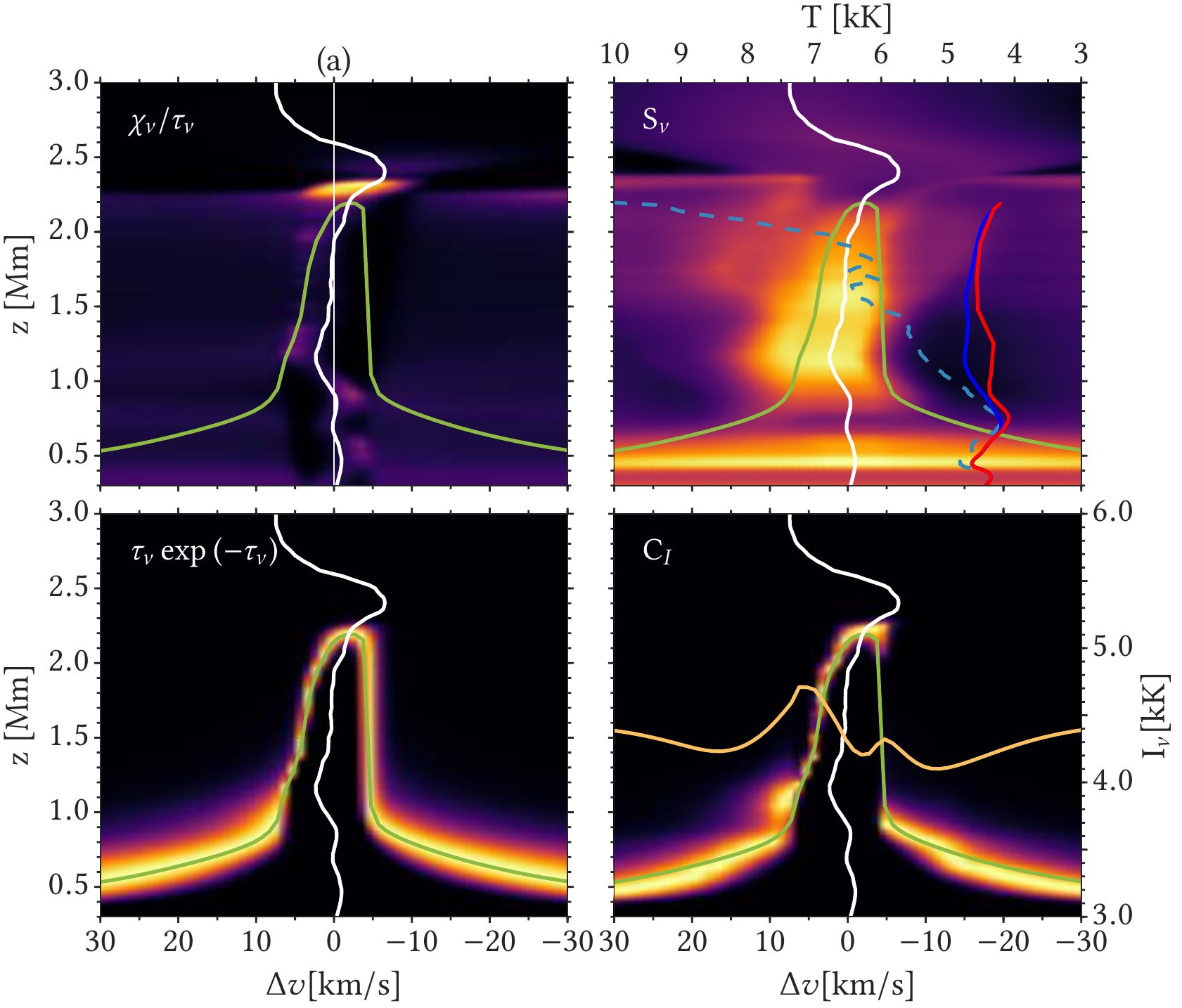}\hfil
    \includegraphics[%
      width = 0.5\textwidth,%
      trim = 2pt 2pt 5pt 3pt,%
      clip = true%
    ]{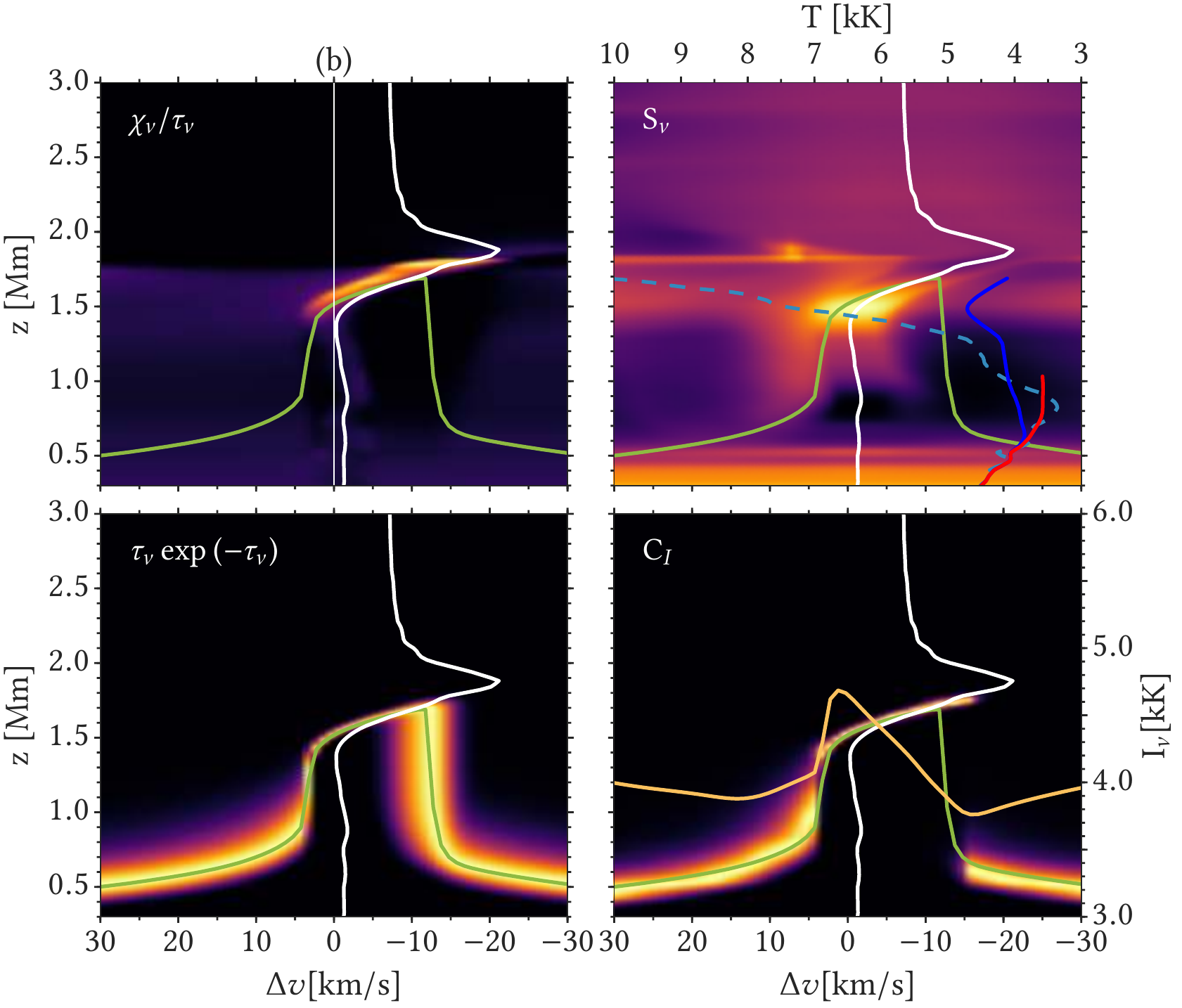}
    \\
    \includegraphics[%
      width = 0.5\textwidth,%
      trim = 2pt 3pt 2pt 3pt,%
      clip = true%
    ]{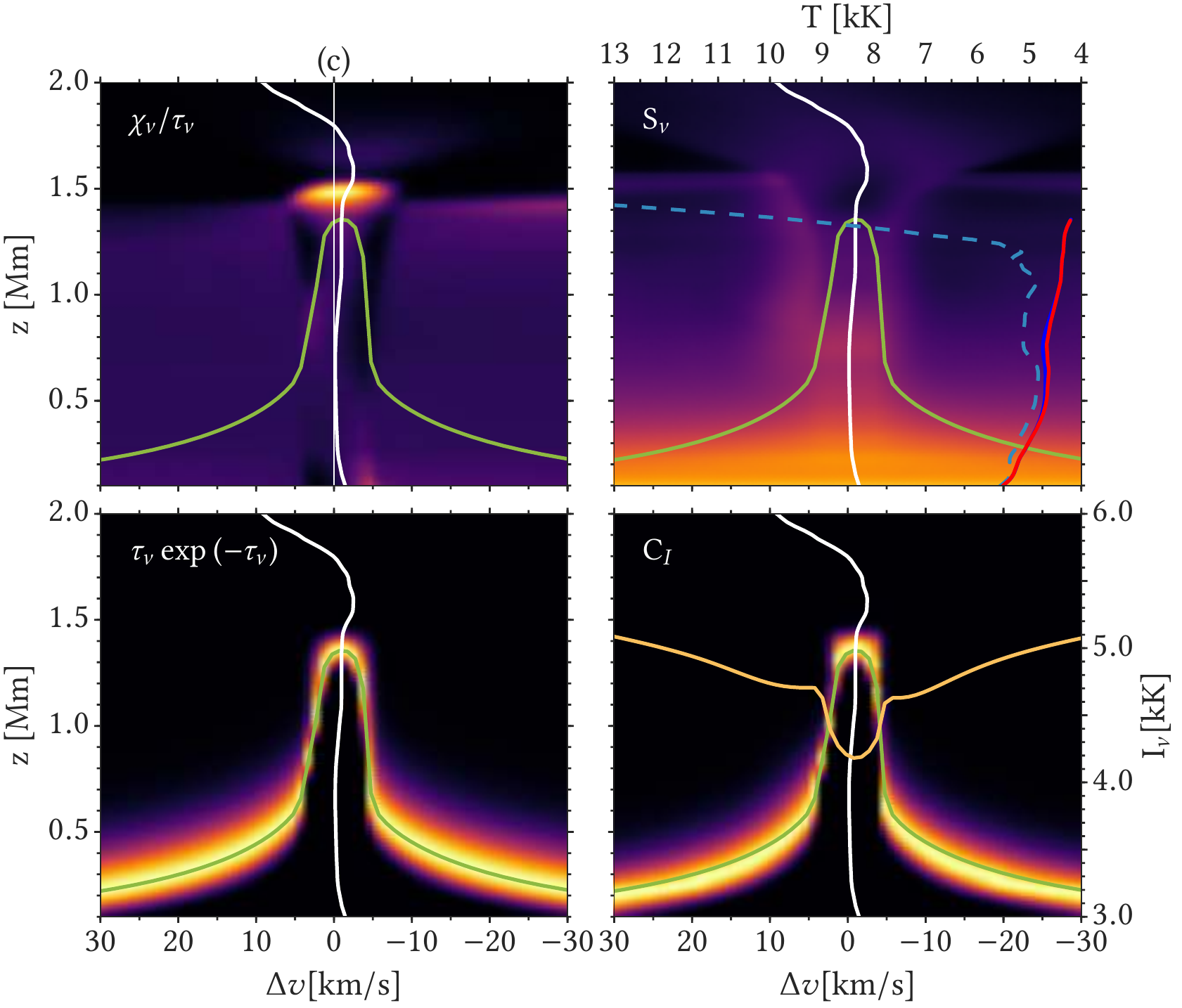}\hfil
    \includegraphics[%
      width = 0.5\textwidth,%
      trim = 2pt 3pt 5pt 3pt,%
      clip = true%
    ]{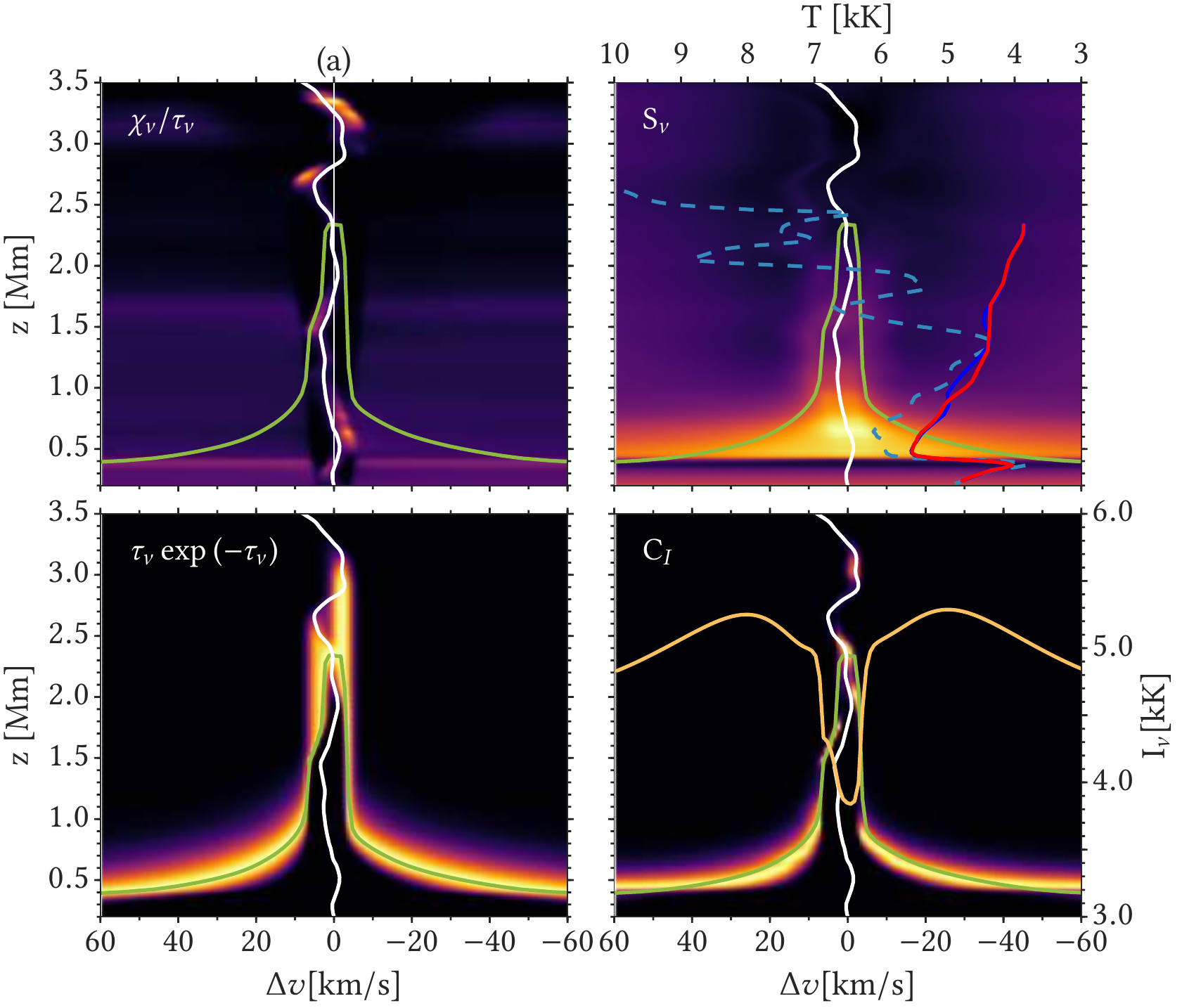}%
  \end{minipage}%
  \caption{%
  Intensity formation breakdown figure for the \ion{Ca}{II}~K line from Model 2.
    Subfigure~a): normal profile with two emission peaks.
    Subfigure~b): profile with a single emission peak.
    Subfigure~c): pure absorption profile.
    Subfigure~d): {profile with a large peak-to-peak separation.}
    Each image (black corresponds to low values, yellow to high values)  shows the quantity specified in
    its top-left corner as function of frequency from line center (in
    Doppler shift units) and simulation height $z$.  Multiplication of
    the first three produces the intensity contribution function in
    the fourth panel. A $\tau_\nu=1$ curve (green) and the
    vertical velocity (white solid, positive is upflow) are
    overplotted in each panel, with a $v_z = 0$ line in the first
    panel for reference.  The upper-right panel also contains the
    {gas temperature} (dashed) and the line source function along the
    $\tau= 1$ curve in blue for the part of the $\tau$ curve blueward
    of its maximum value and red for the part on the red side of the
    maximum $\tau = 1$ height, in temperature units specified along the
    top. The lower-right panel also contains the emergent intensity
    profile (orange), as brightness temperature with the scale along the
    right-hand side.}
  \label{fig:four_panel_1}
\end{figure*}

We discuss the formation of four example K line profiles by using the method of
\citetads{1997ApJ...481..500C} 
{computed in 3D XRD with Model 2.}
We decompose the contribution function to the emergent intensity
\begin{equation} \label{eq:C_I}
  \newcommand*{\dd}{ \ensuremath{ \mathrm{d} } } 
  C_I( \nu, z )
    \equiv
    \dfrac{ \dd I(\nu, z) }{ \dd z }
    =
    \dfrac{ \chi(\nu, z) }{ \tau(\nu, z) } \cdot
    S(\nu, z)\cdot
    \tau(\nu, z) \exp\bigl( -\tau(\nu, z) \bigr)
\end{equation}
into three components, $\chi/\tau$, $S$, and $\tau \exp( -\tau )$, which we plot
side by side in a 2$\times$2 panel diagram showing the dependence on the
frequency $\nu$ along the X-axis and the dependence on the height $z$ along the
Y-axis. On each panel we overplotted optical depth unity as function of frequency $z(\tau_\nu =1)$, as well as the vertical velocity $v_\text{Z}(z)$.

The first component is the ratio of the total {(line plus continuum)} opacity $\chi$, to the optical
depth $\tau$. It is dominant at small optical depths and is sensitive to the line-of-sight velocity gradient.

The second component is the total source function $S$, which is
frequency-dependent for PRD lines.
To emphasize this feature, we plot the source function $S$ in temperature units along the redward and blueward slopes of the $z(\tau_\nu =1)$ curve.
We also show the local gas temperature.

The third component $\tau \exp( -\tau )$ outlines where the optical depth
$\tau$ equals unity.

The contribution function $C_I$ is shown on the last panel together with the
emergent intensity profile $I(\nu)$.

Figure~\ref{fig:four_panel_1} provides 2$\times$2 diagrams for four types of
intensity profiles that we used to classify the features.
A normal profile with two emission peaks (Subfigure~a), a profile with a
single emission peak (Subfigure~b),
a pure absorption profile (Subfigure~c), and a complex profile with many
emission peaks (Subfigure~d).

Subfigure~{(a)} shows the formation of a normal profile with two emission peaks and
K$_\text{2V}$ is stronger than K$_\text{2R}$.
The positive asymmetry is caused by a combination of an upflow at 1.2~Mm and a
downflow at 2.4~Mm and by an enhancement of the source function at 1.2~Mm.
K$_\text{2V}$ is formed at 1.3~Mm with $T_\text{b} = 4.9$~kK, which is roughly
500~K lower than the gas temperature at the same height.
K$_\text{2R}$ is formed at 0.9~Mm and has an upper-chromospheric contribution
from 2.1~Mm.
K$_3$ is well-formed as a central depression formed at the maximum formation height of 2.2~Mm
and its Doppler-shift matches the vertical velocity at this height.

PRD effects add an extra complexity to the analysis.
The line source function is not constant anymore and varies strongly with
wavelength.
The source function starts decoupling from the Planck function already
at $\pm$30~km\,s$^{-1}$ around the line core and then it strongly diverges at
$\pm$12~km\,s$^{-1}$.

Subfigure~{(b)} shows a profile with a single emission peak.
This shape is caused by a strong downflow at 1.5--1.9~Mm in the wake of a shock wave that has passed before.
This profile has no K$_3$, and the maximum formation height is located in the blue
slope of the emission peak, at $\Delta\varv = -10$~km\,s$^{-1}$.
The only emission peak is formed at 1.6~Mm and is identified as K$_\text{2V}$
being on the blue side off the line center, at $\Delta\varv = +2$~km\,s$^{-1}$.
The source function and the Planck function are decoupled from each other at
K$_\text{2V}$.
The difference in the observed brightness and the local gas temperatures is
2.1~kK.
The  source function and the Planck function are almost coupled at
K$_\text{1R}$ and K$_\text{1V}$ located at $\Delta\varv = -13$~km\,s$^{-1}$
and $\Delta\varv = +12$~km\,s$^{-1}$ respectively, so that the brightness
temperature at both minima corresponds to the local temperature.

Subfigure~{(c)} shows a pure absorption profile.
Throughout the entire range of the formation heights, the vertical velocity does
not exceed 2.5~km\,s$^{-1}$.
This makes the formation height profile almost symmetric around the line center and the
shape of the intensity profile is mostly defined by the variation of the
source function with height.
The source function is well-coupled to the Planck function up to 0.6~Mm and they
follow a very flat slope.
These absorption profiles usually appear if the vertical velocity amplitude is
small,  and the chromosphere is cold without a strong temperature increase.


{
Subfigure~{(d)} shows a profile with large peak-to-peak separation.
Both emission peaks, at $\pm$27~km\,s$^{-1}$, are caused by a deep chromospheric temperature rise 
at $z=0.35$~Mm. The emission peaks are symmetric because the vertical velocity is only 1~km\,s$^{-1}$ at the formation height. The source-function decouples from the Planck function at 0.4~Mm and decreases towards the maximum formation height at $2.7$ Mm, forming the central core. 
}


\section{Comparison between observations and simulations}
\label{sec:comparison_obs}

We compared our synthetic data with the SST/CHROMIS observations
made in the K line.
We also tried to reproduce some general properties of the K line in the solar
spectrum using older data.
As the H and K lines share many common properties and are formed in the
chromosphere in practically the same way, we expect our conclusions to be
similar for them both.

The primary reason to prefer the K line to the H line is that the former has a factor two higher opacity 
and therefore is formed slightly higher in the chromosphere. This allows to
probe the largest height range in the atmosphere.
In addition, the H line typically has less pronounced  H$_\mathrm{2V}$ and
H$_\mathrm{2R}$ emission peaks.

{Another practical reason is that the H line is blended with the H-$\epsilon$
line of \ion{H}{I} at 3\,970.075~\AA, that is, in the red wing just next to the
H$_\text{1R}$ feature.}

\subsection{Spatially-averaged \ion{Ca}{II}~K spectrum} \label{subsec:spat_aver}

\begin{figure}
  \begin{minipage}{\columnwidth}
    \includegraphics[%
      width = \columnwidth,%
      trim  = 0pt 2pt 0pt 2pt,%
      clip  = true%
    ]{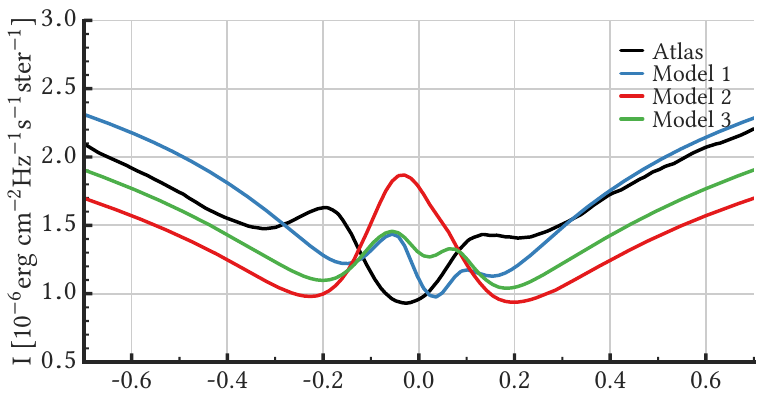}
    \\
    \includegraphics[%
      width = \columnwidth,%
      trim  = 0pt 2pt 0pt 2pt,%
      clip  = true%
    ]{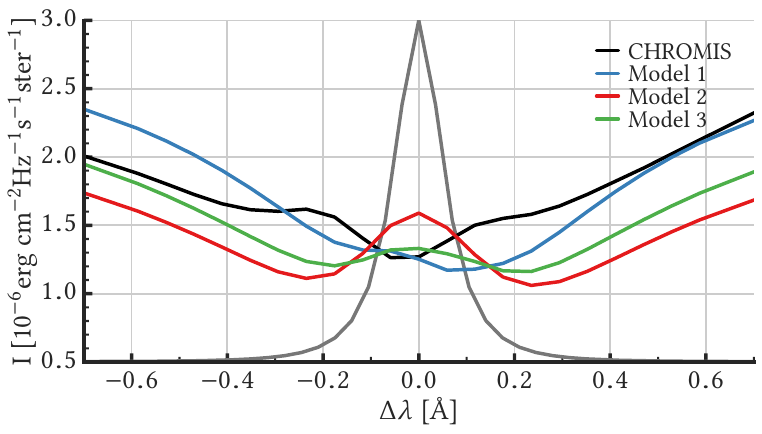}%
  \end{minipage}%
  \caption{%
    Spatially-averaged intensity profiles of the \ion{Ca}{II}~K line at
    $\mu = 1$.
    \emph{Upper panel:} undegraded synthetic profiles for Model~1, 2, and 3
    (blue, red, and green) are compared with the Hamburg atlas profile (black).
    \emph{Lower panel:} degraded synthetic profiles (same notation) are compared
    with the SST/CHROMIS profile from our observations (black). The grey curve indicates the assumed CHROMIS spectral transmission profile.
  }
\label{fig:atlas_solar}
\end{figure}

We compare spatially-averaged intensity profiles of the K line from our
simulations computed with 3D XRD with spatially-averaged profiles from our observations and high-resolution
profiles from the Hamburg {quiet-Sun} disk-center intensity atlas
\citepads{1984SoPh...90..205N,  
          1999SoPh..184..421N}. 

Figure~\ref{fig:atlas_solar} (\emph{top}) relates undegraded profiles from our
calculations to the atlas profile measured with
$\lambda/\delta\lambda = 4.5\cdot 10^5$ resolution.
The atlas profile shows distinct K$_\text{1V}$, K$_\text{2V}$, K$_3$,
K$_\text{2R}$, and K$_\text{1R}$ features, with K$_\text{2V}$ stronger than
K$_\text{2R}$ and separated by 0.3~\AA, and with K$_3$ shifted by $-$0.03~\AA\
off the nominal  line center.
The two separated emission peaks are reproduced by Model~1 and Model~3, while
they are blended together in Model~2 and separate only towards the limb (see
Fig.\ \ref{fig:center_2_limb} and Section~\ref{sec:c2l-variation}).
The peak separation in all three models is less than one half of the peak
separation in the atlas profile.
The same is true for the K$_\text{1V}$-to-K$_\text{1R}$ distance, which is the
biggest, but still insufficient, in Model~2.
The peak asymmetry is correct in Model~3, stronger in Model~2, and too big in
Model~1.
Peak intensities are too low in Model~1 and Model~3 and are too high in
Model~2.
The opposite is true for the line core intensities.
The K$_3$ core is red shifted in Model~1 and Model~3 and might be at the right
position in Model~2 as the blended peaks are slightly blue shifted off the line
center.
The K$_1$ features as well as the inner wings have a lower intensity in the simulations than in the atlas, except
for the outer wings in Model~1, which are brighter than in the atlas profile.

Figure~\ref{fig:atlas_solar} (\emph{bottom}) relates the spatially-averaged profile observed with
SST/CHROMIS with the simulated line profiles degraded to CHROMIS spectral resolution.

In the observed profile, we can still recognize all the features although K$_\text{2R}$ is only a small bump in the red flank of the line and not a {clear} local maximum.
The peak asymmetry and the K$_2$ intensities are reduced, while the K$_3$
intensity is increased.
The wavelength positions as well as the corresponding separations of the
features are almost unaffected.
The spatial and spectral resolution of the instrument smooth out small spectral
features in the synthetic data so that the K$_3$ core disappears and the K$_2$
features cannot be resolved as two separate peaks.
The K$_1$ and inner wing intensities remain roughly the same while the K$_2$ intensities
are reduced.

None of the models reproduce both the full-resolution and the
degraded K line profiles. 
The models appear either too cold or too hot in the upper photosphere, and they are too
cold around the temperature minimum
and are either too cold or too hot in the middle chromosphere, where non-thermal
broadening is not strong enough in the simulations.

\subsection{Center-to-limb variation}
\label{sec:c2l-variation}

\begin{figure}
  \includegraphics[%
    width = \columnwidth,%
    trim  = 0pt 2pt 0pt 1pt,%
    clip  = true%
  ]{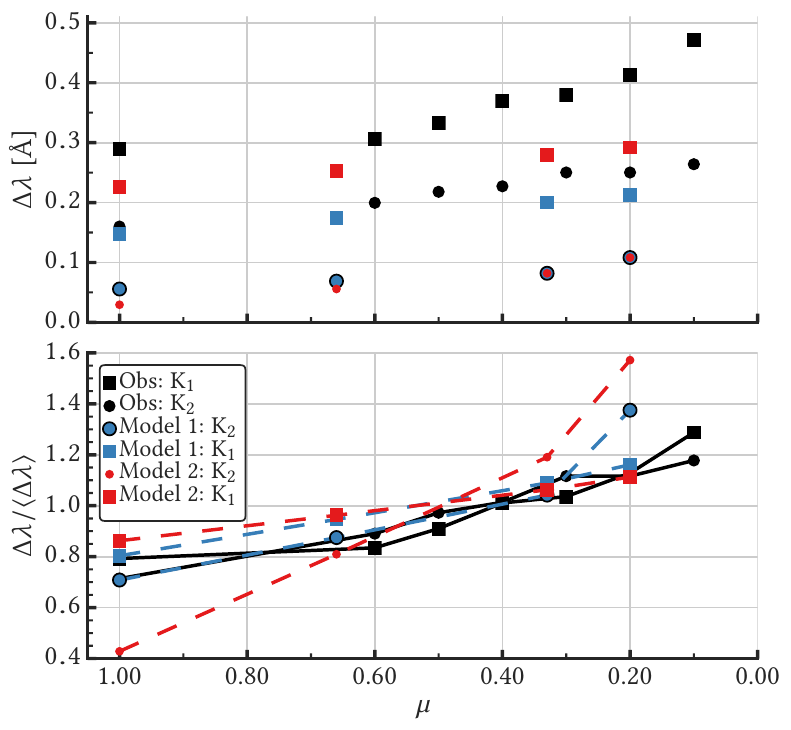}%
  \caption{%
    Center-to-limb variation of the K$_\text{1V}$ and the K$_\text{2V}$
    distances off the line center observed by
    \citetads{1968SoPh....3..164Z} 
    and
    \citetads{1975ApJ...199..724S} 
    (black markers), and computed for Model~1 (blue markers) and Model~2 (red
    markers) for the spatially-averaged spectrum of \ion{Ca}{II}~K.
    \emph{Upper panel}: absolute distances $\Delta\lambda$ to the features at
    different $\mu$ angles.
    \emph{Lower panel}: trends of $\Delta\lambda$ variations normalized to the
    mean $\langle\Delta\lambda\rangle$ in each group.}
  \label{fig:c2l-variation}
\end{figure}

Center-to-limb observations of the solar \CaII~K spectrum show two effects
\citepads[see, e.g.,]{1966ApNr...10..101E, 
                      1968SoPh....3..164Z} 
%
First, all their intensities undergo limb-darkening.
Second, the K$_1$ and K$_2$ separations increase towards the limb.
First shown by
\citetads{1975ApJ...199..724S} 
and later confirmed by more accurate modeling by
\citetads{1989A&A...213..360U}, 
the H and K lines must be treated in PRD as modeling assuming CRD cannot reproduce any of the
center-to-limb effects.
However, it is not possible to accurately model both effects using the same 1D
model atmosphere
\citepads{1975ApJ...199..724S}. 

We tested whether we can reproduce both center-to-limb effects in Model~1 and
Model~2.
We computed spatially-averaged K-line intensities at $\mu = 1.0$, 0.66, 0.33,
and 0.2 in the most accurate 3D~XRD treatment.
{We computed the intensity output for two azimuths: 0 and 90 degrees, that
is, along the X-axis and the Y-axis and four different latitude directions having
$\mu_\text{Z}$ = 1, 0.66, 0.33, and 0.2. For each latitude we average over the two azimuths.}
For comparison, we adopted the observations taken with the Sacramento Peak
Observatory spectrograph
\citepads{1968SoPh....3..164Z, 
          1975ApJ...199..724S} 
from the disk center towards the south pole of the Sun, with the slit aligned in
the North-South direction and an exposure time of 30~s.

Figure~\ref{fig:center_2_limb} illustrates the first center-to-limb effect in
the K line.
All features as well as the inner wings of the K line undergo a limb darkening
in both model atmospheres, that is, their intensities steadily decrease towards
the limb.
We note that the emission peaks at K$_\text{2V}$ and K$_\text{2R}$ as well as
the outer minima at K$_\text{1V}$ and K$_\text{1R}$ become more separated
towards the limb.

Figure~\ref{fig:c2l-variation} relates the observed and synthesized distances of
K$_\text{1V}$ and K$_\text{2V}$ off the line center.
In this figure, we show variations of the absolute values (upper panel) as well
as slopes of their trends (lower panel).

We note the same problem discussed above that the calculated K$_1$ separations are
much smaller {than} the observed ones.
Model~1 produces less than 50\% and Model~2 produces 70--80\% of the
observed widths.
The K$_\text{1V}$ features show similar trends in both models.
These trends are  flatter than the observed one.
The K$_\text{2V}$ features show very steep trends in both models, while the
observed trend is a bit less steep than the observed K$_\text{1V}$ trend.

Although both model atmospheres do not reproduce correct separations between the
corresponding features, they do reproduce the observed trends in center-to-limb behaviour.
In a certain sense, Model~1 fits better as it does not show extreme the trends of
Model~2.

\subsection{Statistics of the \CaII~K line parameters}
\label{sec:obs-line-parameters}

\begin{figure}
    \includegraphics[%
    width = \columnwidth,%
     clip  = true%
  ]{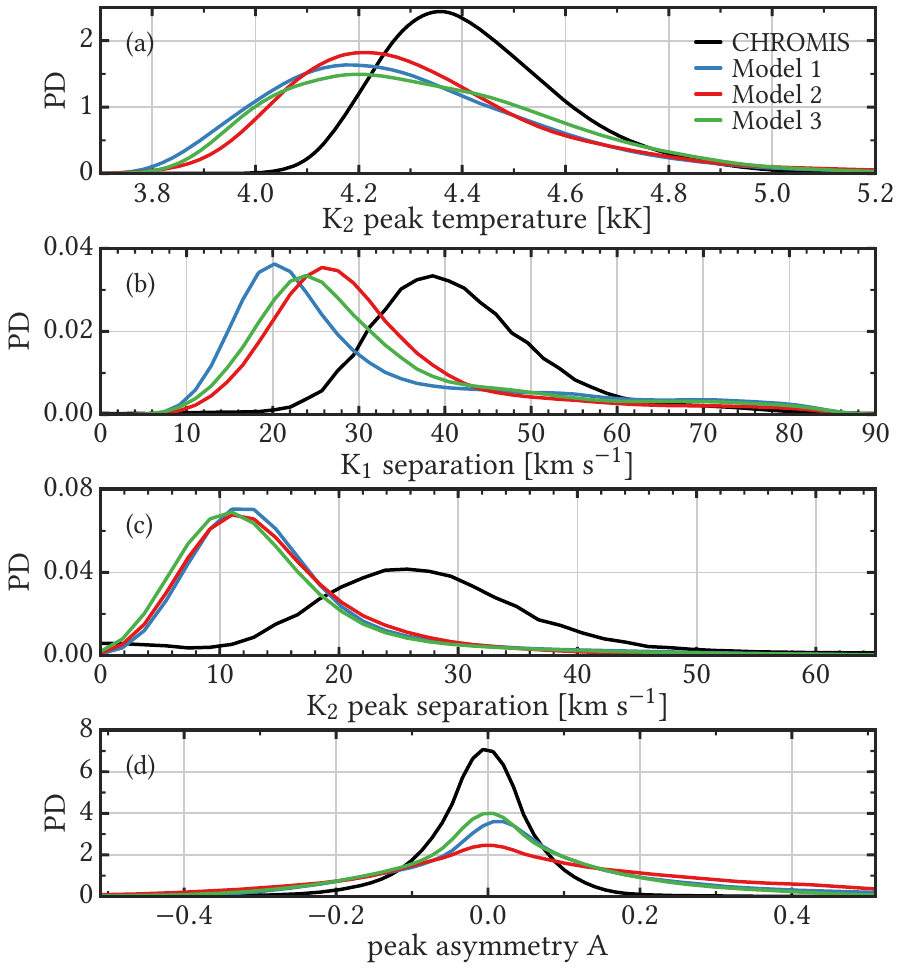}%
  \caption{%
    From top to bottom: distributions of the \KtwoV\ and \KtwoR\ radiation temperature, the \Kone\  separation,
   the \Ktwo\  peak separation, and the peak asymmetry in the SST/CHROMIS
    observations (black) compared to the simulations in Model~1, 2, and 3 (blue,
    red, and green) {for $\mu=1$}.}
  \label{fig:obs3}
\end{figure}

Using the observed CHROMIS dataset and our synthetic data sets, we investigated the distributions
of four observable K-line parameters:
1) the brightness temperature at the \Ktwo\ emission peaks $T_\text{b}(\text{K}_2)$;
2) the wavelength separation between the \Kone\ minima;
3) the wavelength separation between the \Ktwo\ maxima;
4) the peak asymmetry
\begin{equation} \label{eq:peak-asymmetry}
  A =
    \dfrac{
      I(\text{K}_\text{2V}) - I(\text{K}_\text{2R})
    }{
      I(\text{K}_\text{2V}) + I(\text{K}_\text{2R}).
    }
\end{equation}

We spatially degraded the synthetic dataset as in Sec.~\ref{sec:datadegradation}. The simulated emission peaks are roughly a factor 2.1 narrower than the observations, as shown in Fig.~\ref{fig:atlas_solar}. Degrading with the spectral resolution of CHROMIS (120~m\AA) would lead to unrealistically many single-peaked profiles and too-low \Ktwo\ intensities and too high \Kone\ intensities. We therefore smeared with a Gaussian of $120/2.1$~m\AA$ = 57$~m\AA, where the factor 2.1 comes from the difference in emission peak width. This lower value is chosen so that the simulated profiles are smoothed, just as in the observations, but not so much that the emission peaks blend together. We note that this comparison is somewhat unfair. However, given the difference in the width of the central emission peaks, it allows a reasonable comparison of the distribution of the profile parameters.


Figure~\ref{fig:obs3} shows distributions of all four parameters in the
observations and in the simulations.
The models predict a too low \Ktwo\ brightness temperature compared to the observations, most likely caused by a too low temperature in the middle chromosphere in the simulations. The median brightness temperature in the simulations is 4.2~kK, while the observed one is 4.4~kK. The straylight contamination in the observations can affect the $T_\text{B}(\text{K}_2)$ distribution by decreasing its dynamical range of temperatures, so the real discrepancy between observations and simulations might be smaller than implied by the distributions.

The mean \Kone\ separation is smaller in the models than is observed, and their distribution has a long asymmetric tail towards high values, while the observed distribution is more symmetric.
Likewise, the simulated \Ktwo\ separations are on average lower and show a tail in the distribution towards high values. The observed distribution is more symmetric.
The \Kone\ and \Ktwo\ separation distributions are not sensitive to straylight contamination and the disagreement between the models and the observations
 means that physical processes that produces non-thermal
line broadening are missing or not sufficiently strong in the models.
Previously, similar effects have been reported for Model~1 by
\citetads{2013ApJ...772...90L} 
for the \MgII~h and k lines and by
\citetads{2015ApJ...811...80R} 
for the \ion{C}{II} 1335~\AA\ triplet.

Finally, the peak asymmetry shown in the bottom panel of Fig.~\ref{fig:obs3} can be used to assess strong velocity gradients caused by shock waves traveling
upwards in the chromosphere
\citepads{1992ApJ...397L..59C,  
          1997ApJ...481..500C}. 
It is known from observations
\citepads[e.g.,][]{1974SoPh...37...85G,  
          1983ApJ...272..355C,  
          2008A&A...484..503R}, 
that K$_\text{2V}$ is usually stronger than K$_\text{2R}$ meaning that the peak
asymmetry is slightly negative.
All three distributions are centered around zero showing mean values of $-0.01$
(observations), and 0.01--0.04 (simulations).
The observed distribution is negatively skewed with a bit longer tail of
negative values and it goes from $-0.3$ to 0.2.
The synthetic distributions are positively skewed with a fairly long tail of
positive values and they go from $-0.3$ to 0.5.
The synthetic distributions are wider. The closest, although not good, agreement with the observations appears
in Model~3.
The observed peak asymmetry distribution is not very sensitive to straylight. The model distributions might be influenced by the fact that we use a single snapshot from each simulation run. The mean and the skewness of the distribution might depend on the phases of the global box oscillations which are present in the simulations
\citepads[see Fig.~8 of][]{2016A&A...585A...4C}.
%

\subsection{Images in the \CaII~K line}
\label{sec:imaging-in-K-line}

\begin{figure*}%
    \includegraphics[%
    width = \textwidth,%
    trim  = 0pt 1pt 2pt 1pt,%
    clip  = true%
  ]{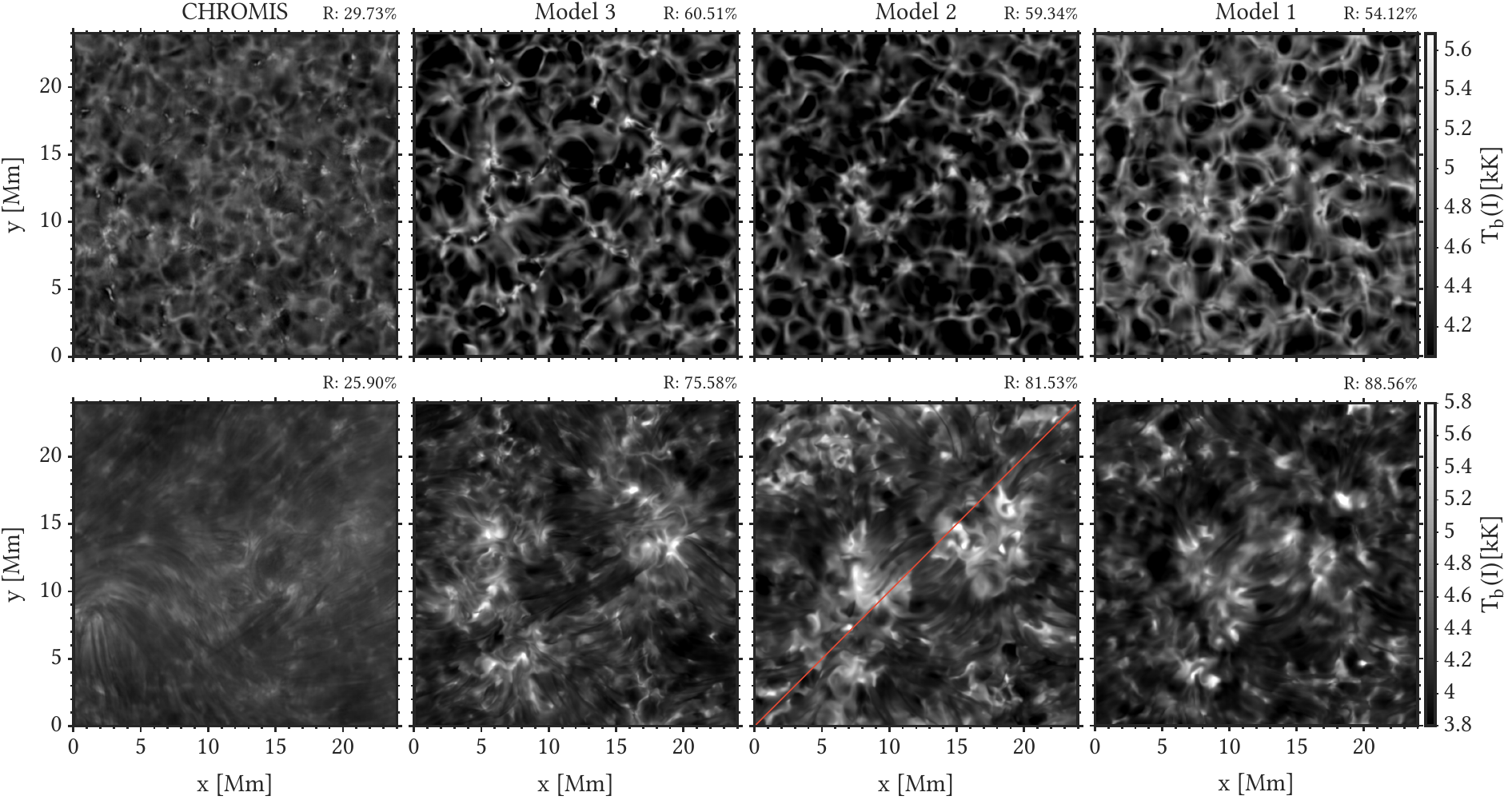}%
  \caption{%
    Observed and simulated \ion{Ca}{II}~K images of the quiet Sun at the disk
    center ($\mu = 1$).
    \emph{Upper row}: line wing at $\Delta\lambda = -0.528$~\AA.
    \emph{Lower row}: line center at $\Delta\lambda = 0$~\AA.
    SST/CHROMIS images (\emph{first column}) are compared with simulated
    images using Model~3, 2, and 1 (\emph{second, third, and fourth
    columns}).
    Images are given in brightness temperature units with scales common for
    each row, specified at the right.
    The r.m.s.\ contrast $R$ is provided in the upper right corner above each
    image.
    The red line indicates the slice through Model~2 shown in
    Fig.\ \ref{fig:figatmos}.
  }
  \label{fig:obs2}
\end{figure*}

We investigate whether Bifrost simulations reproduce similar chromospheric
structures as we see in observed \CaII~K images.
Figure~\ref{fig:obs2} relates line-wing and line-core images observed with
SST/CHROMIS to similar images synthesized using the three Bifrost model
atmospheres.  Again, we spatially smeared the simulated images to SST/CHROMIS resolution, but  smeared spectrally with a Gaussian with FWHM of 57~m\AA. Smearing the simulated images with the full CHROMIS profile leads to too much mixing of low-chromospheric signal in the line-core images, and would not allow us to compare the fibril structure formed in the mid and upper chromosphere. We stress again that this is not a fair comparison of the imagery, and the small peak separation in the synthetic profiles is a clear indication that the chromosphere is not yet modeled correctly in the Bifrost models.


The line-wing images were observed and simulated close to the K$_\mathrm{1V}$
features, so they sample the upper photosphere just below the temperature minimum. We see regular brightness patterns in all four cases. These patterns come from reversed granulation, acoustic wave fronts and magnetic features.
The Bifrost models show a brightness
pattern with similar spatial scales in the most quiet regions (in the corners) as in the observations but with larger spatial scales between the two polarities in the central part of the field of view. This region has flux-emergence in the simulations and also larger size granules. All the simulations show higher contrast compared to the observations.
On average, Model~2 and Model~3 are darker and Model~1 is brighter than the
observations.
The typical size of magnetic field concentrations in intergranular lanes is also
larger in  the synthetic images: they appear as bright tiny dots of 4.8--5.1~kK in
the observations while in the simulations they look like bright, diffuse, and
elongated spots of 5.0--5.2~kK.
Cold granular patches can be quite dark in the simulations, with values below
4.0~kK, while in the observations they are somewhat higher, typically 4.1--4.2~kK.

The line core images in Figure~\ref{fig:obs2}  show images at the nominal line
center.
The observations show a rather diffuse pattern of reversed granulation and sub-canopy shock waves away from magnetic field concentrations, with superimposed fibrilar structure that appears semi-transparent. The fibrils are  thin (less than 0.3~Mm wide), are typically 5--10~Mm long, and many originate at
bright patches of magnetic concentrations seen in the observed line-wing image. The observed fibrils cover the whole field of view, and they are  typically only slightly curved.

The synthetic images have a higher contrast than the observations. The areas above the photospheric magnetic elements are bright. The 96-km resolution Model~1 shows only a few fibrils, Model~2 with 48~km resolution shows more, and Model~3 with 31~km resolution shows most. The fine fibrils in Model 3 are quite reminiscent of the observed ones. Still all models show too few fibrils and too strong visibility of the underlying shocks and reversed granulation.
The numerical simulations have a horizontal grid spacing of 31 or 48~km. The smallest structures that can be formed in the simulation are roughly four times the grid spacing, and are thus of comparable size or larger than the spatial resolution of CHROMIS/SST. Simulations with higher spatial resolution are thus required to resolve the smallest observable scales.

The hottest network structures are brighter than 5~kK in the simulations and
have only 4.7--4.9~kK in the observations.
The coldest internetwork patches are colder than 4.1~kK in the simulations and
are 4.1--4.2~kK in the observations.

We conclude that visually Model~3 shows more fine details than Model~1 or
Model~2, and the structures in Model~3 are more similar to the observed chromospheric structures than those in Models 1 and 2.

In terms of the root-mean-square contrast of the intensity, all Bifrost
models produce a factor of 1.8--2.0 higher contrast in the line wing and a
factor of 2.9--3.4 higher contrast in the line core in comparison with the 
observations. This discrepancy is at least partially caused by the straylight contamination, which we
did not correct for in the reduction procedure.

\section{Diagnostic potential of the H and K lines}
\label{sec:result_diagnostic}

We use undegraded synthetic K-line spectrograms  to demonstrate which
properties of the line profile are useful for diagnosing the chromosphere.
We discuss results for the K line only, because the H line is formed in the same
way.
In some sections we employ only one model atmosphere out of three because the
other models show similar results.

\subsection{Formation heights of the profile features}

\begin{figure*}
  \includegraphics[width = \textwidth]{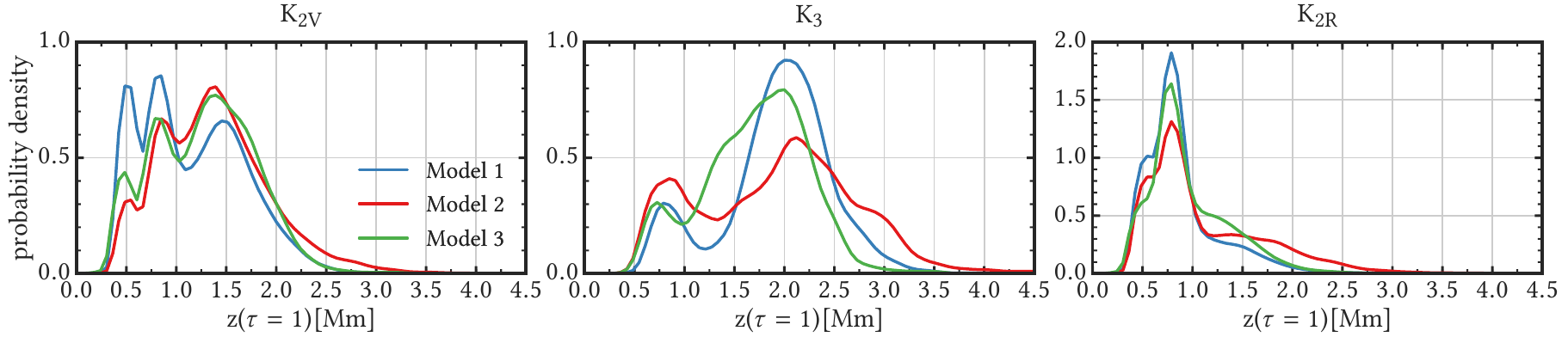}
  \caption{%
    Formation heights of K$_\text{2V}$ (\emph{left}), K$_3$ (\emph{middle}),
    and K$_\text{2R}$ (\emph{right}). {The zero-point is defined as the average height where the optical depth at 5000 \AA\ is unity.}
    Distributions show the probability density for heights where the optical
    depth equals unity in Model~1 (blue), Model~2 (red), and Model~3 (green).%
  }
  \label{fig:fig3_height}
\end{figure*}

We identify the K-line profile features in all three model atmospheres and
measure corresponding formation heights. {The zero-point for the formation height is defined as the average height where the optical depth at 5000 \AA\ is unity.}
Using the Eddington-Barbier approximation at $\mu = 1$, we define the formation
height $z$ at a given frequency $\nu$ as the height where the optical depth equals unity, $\tau_\nu = 1$.
Figure~\ref{fig:fig3_height} shows the obtained distributions of $z(\tau_\nu = 1)$ at the frequency of the \Ktwo\ and \Kthree\ features.

The \Kthree\ feature is formed in the widest range of heights at 0.5--4.0~Mm.
The side lobe on the left side of the distribution near 0.5--1.0~Mm consists of pixels where the feature-finding algorithm failed.
Less than half of them are when K$_3$ is mistaken for K$_1$ around a single
emission peak (see Sect.~\ref{sec:line_features}).
The rest are when K$_3$ is mistaken for some local minimum in a complex profile
with many emission peaks. Such local minima are formed between the upper photosphere and the lower
chromosphere.
On average, K$_3$ forms at 1.9{$\pm 0.6$}~Mm.

The \Ktwo\ emission peaks are on average formed below \Kthree, at 0.3--3.0~Mm.
On average, K$_\text{2V}$ forms at 1.3{$\pm 0.5$}~Mm and
K$_\text{2R}$ forms at 1.0{$\pm 0.5$}~Mm.
Model~2 shows the biggest mean formation heights, 1.4~Mm for K$_\text{2V}$ and
1.1~Mm for K$_\text{2R}$.

The K$_\text{2V}$ distributions show three maxima in all three models.

The leftmost side maximum at 0.5~Mm is obtained from profiles with a single
emission peak.
The emission peak is produced in the lower chromosphere by strong downflows
following the upward passage of shock waves.

The next peak mode at 0.8~Mm is caused by complex profiles with many
emission peaks, where the standard classification cannot by applied.

The principal peak at 1.3--1.4~Mm is obtained from standard profiles with two
emission peaks.
It is strong for K$_\text{2V}$ and is weak for K$_\text{2R}$ as the red emission
peak is often difficult to measure in normal profiles because it is weak or appears as a slight bump in the line profile but not a local maximum.

Features of the H line are formed below the corresponding features of the K
line.
On average, H$_3$ forms 150~km below K$_3$, H$_\text{2V}$ forms 150~km
below K$_\text{2V}$, and H$_\text{2R}$ forms 100~km below K$_\text{2R}$.

\subsection{The K$_3$ minimum}

\begin{figure*}
  \includegraphics[%
    width = \textwidth,%
    trim  = 2pt 4pt 2pt 6pt,%
    clip  = true%
  ]{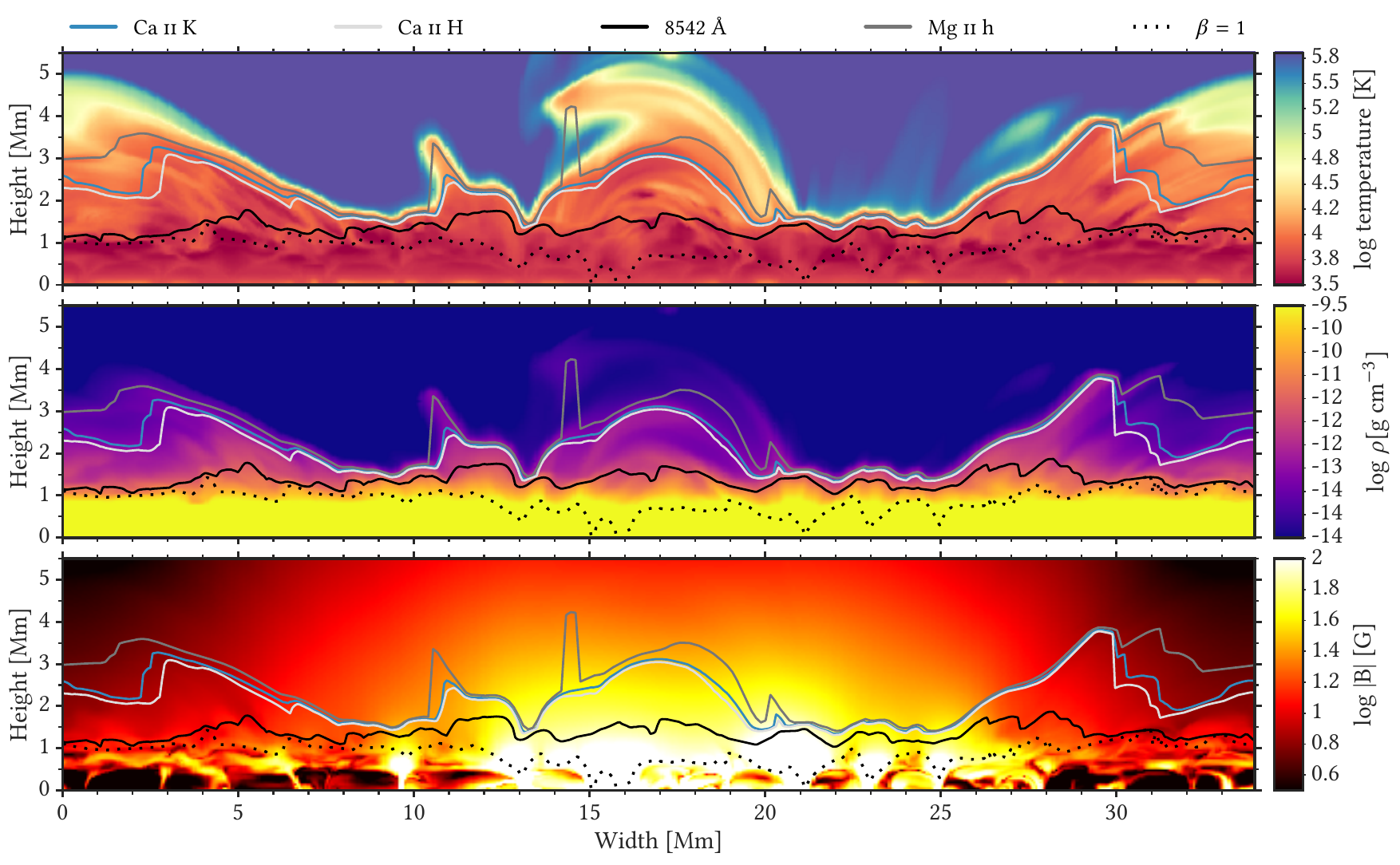}%
  \caption{%
    Maximum formation heights,  {max $z(\tau_\nu=1)$} , of the \ion{Ca}{II}~K (blue), the
    \ion{Ca}{II}~H (white), the infrared \ion{Ca}{II}~triplet 8\,542~\AA\ (black solid), and the \ion{Mg}{II}~h (gray) lines in Model~2, sliced vertically
    along the main diagonal,  { where max $z(\tau_\nu=1)$ is taken over all wavelength positions of the respective spectral line profile}. {The zero-point is defined as the average height where the optical depth at 5000 \AA\ is unity.}
    \emph{Upper panel}: gas temperature.
    \emph{Middle panel}: mass density.
    \emph{Lower panel}: magnetic field strength. The plasma $\beta$ parameter is less than unity above and bigger than unity
    below the dotted line.%
  }
  \label{fig:figatmos}
\end{figure*}
The cores of the K and H lines are the most interesting features as they are the
most highly formed parts of the line profiles.
At K$_3$ and H$_3$ one can observe the middle-upper chromosphere, still below
the \ion{Mg}{II} h and k or the \ion{H}{I} Ly-$\alpha$ lines in the UV, but
above the other visible, the \ion{Ca}{II}~T or the \ion{H}{I} H-$\alpha,\beta$
lines.
Figure~\ref{fig:figatmos} illustrates this on the Model~2 slice, marked in
Fig.~\ref{fig:obs2}.

All strong chromospheric line cores are formed at heights where the magnetic
pressure $P_\text{B}$ dominates the gas pressure $P_\text{gas}$ (this is
illustrated by the plasma $\beta = P_\text{gas} / P_\text{B} = 1$ curve).
Owing to a factor $\sim~17$ higher opacity, the \ion{Mg}{II} h and k lines
always form above the \ion{Ca}{II} H and K lines.
This difference in formation heights goes from 40~km to 1\,900~km depending on
the density variation in the atmosphere.
At the same time, the H line always forms below the K line, on average 120~km.
Both the H and K line cores form around 2\,200~km in this slice.
The 8542~\AA\ line forms much below, around 1\,400~km.

At network regions with strong vertical magnetic fields, the transition region
lies much lower and the \ion{Ca}{II} and the \ion{Mg}{II} lines are formed very close to each other
just below the transition region  (for example between a width of 21~Mm and 25~Mm
in Fig.\ \ref{fig:figatmos}).
In the internetwork magnetic fields are more horizontal and the transition region
lies much higher on many density loops, which outline the magnetic field direction.
There, the \ion{Ca}{II} and the \ion{Mg}{II} lines are formed much higher with
a much bigger spacing in between their formation heights (for example between a width of 16--20~Mm).
If we define the transition region as the height where the temperature goes
above 30~kK, then the K line forms, on average in all three models, 1.25~Mm
below the transition region.
We did not find any correlation between the K line intensity and the height of
the transition region.

\begin{figure*}
  \includegraphics[%
    width = \textwidth,%
    trim  = 2pt 3pt 2pt 3pt,%
    clip  = true%
  ]{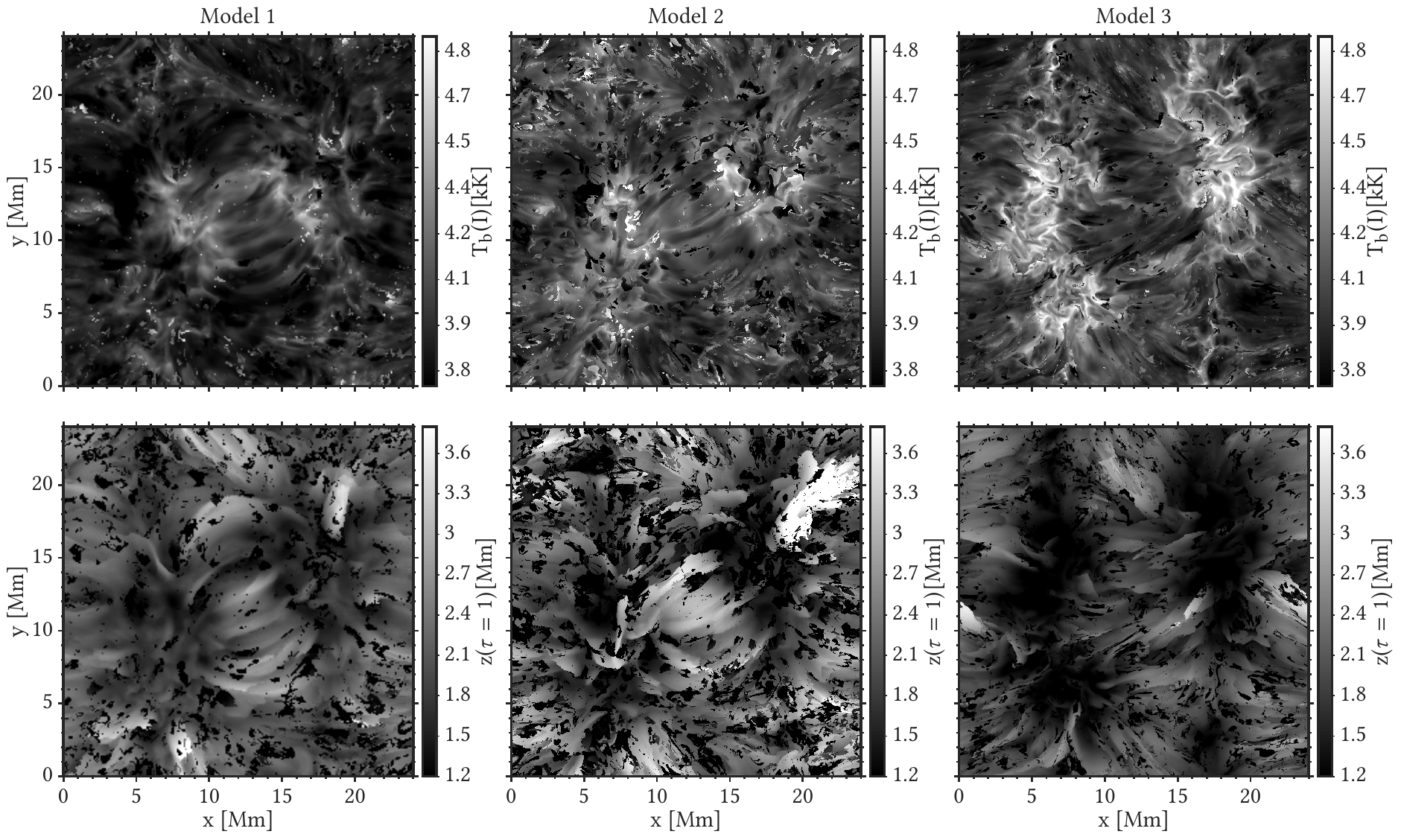}%
  \caption{%
    The emergent intensity expressed as brightness temperature $T_\text{b}$
    (\emph{upper row}) and the formation height $z(\tau = 1)$
    (\emph{bottom row}) at K$_3$ for Model~1, 2, and 3
    (\emph{left, center, and right column}) computed at $\mu = 1$.%
  }
  \label{fig:k3_image}
\end{figure*}

We want to illustrate a common feature of strong chromospheric lines.
For each model, we composed two images in the XY-plane made of intensities and
corresponding formation heights at K$_3$ given in Fig.\ \ref{fig:k3_image}.
In each image there are many  {white sprinkles in the upper panel and black sprinkles in the lower panel that are caused by misfits of the K$_3$ feature.}

The effect of increasing horizontal resolution is visible from Model~1, which
has very diffuse and unsharp structures, to Model~3, which has many small-scale
sharp elements.
The feature we would like to point attention at is the anti-correlation of the
observed intensity with the formation height at K$_3$.
In other words, bright radiation of the network is formed much below dim
radiation of the internetwork formed higher up.

\begin{figure*}
  \sidecaption
  \begin{minipage}[b]{120mm}
    \includegraphics[%
      width = 60mm,%
      trim = 1pt 1pt 1pt 1pt,%
      clip = true%
    ]{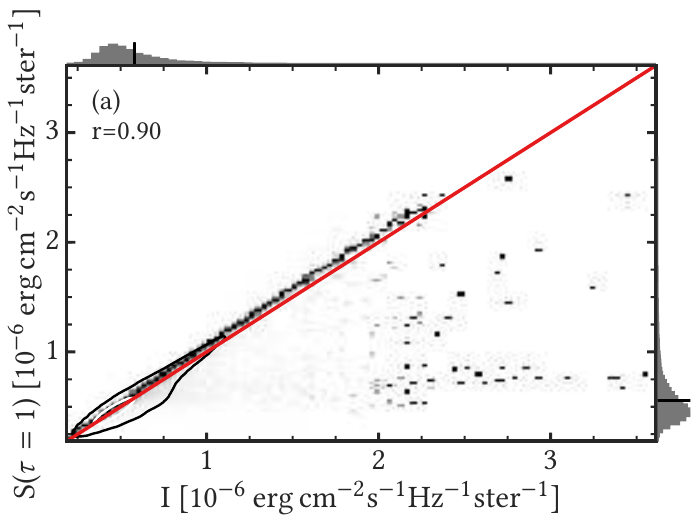}\hfil
    \includegraphics[%
      width = 60mm,%
      trim = 0pt 1pt 2pt -10pt,%
      clip = true%
    ]{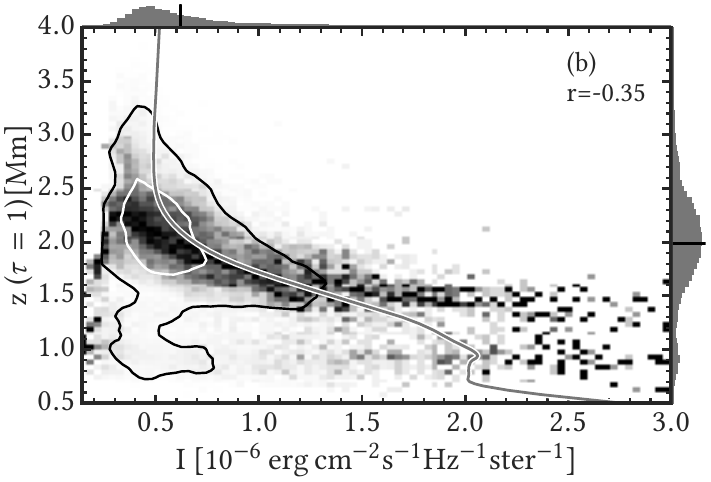}%
  \end{minipage}%
  \caption{%
    Formation properties of \CaII~K$_3$ from Model 1.
    Panel~a): Joint-PDF of the emergent \Kthree\ intensity and the source function at $\tau=1$ at the wavelength of \Kthree.
     Panel~b): Joint-PDF of the emergent \Kthree\ intensity and the $z(\tau=1)$ at the wavelength of \Kthree.
    The grey curve in panel b shows the horizontally-averaged profile-averaged mean intensity $\bar{J}^\varphi\!$.
    The panels follow the same format as Fig.~\ref{fig:fig1_1dprd_vs_3dprd}.%
  }
\label{fig:eddington_b}
\end{figure*}
Panel~a) of Fig.\ \ref{fig:eddington_b} shows the validity of the
Eddington-Barbier relation at $\mu = 1$ for the K$_3$ emergent intensity,
\begin{equation} \label{eq:EB-relation}
  I(\lambda_3, \mu = 1) = S\bigl( \lambda_3, z = z(\tau = 1) \bigr)
\end{equation}

where $\tau=1$ is at the wavelength position of the K$_3$ feature.

As the K line is a strongly scattering one with the photon destruction
probability $\epsilon \approx 10^{-4}$ and as CRD is approximately valid at the
line core, then the line source function at K$_3$ is mostly equal to the angle-
and profile-averaged intensity $\bar{J}^\varphi$.
This is correct for $z(\tau = 1) > 1.3$~Mm as can be seen in panel~b) of
Fig.\ \ref{fig:eddington_b}.
Below $z(\tau = 1) = 1.3$~Mm, the line source function becomes more coupled to
the local Planck function.
The mean intensity of scattered radiation is decreasing with height, that is why
we observe an anti-correlation of $T_\text{b}$ with $z(\tau = 1)$ in
Fig.\ \ref{fig:k3_image}.

\subsection{Diagnostic properties of the H and K lines}

Following
\citetads{2013ApJ...772...90L} 
we investigated what kind of diagnostic the H and K lines can provide for the
chromosphere.
We studied how intensities, wavelength positions, and other derived properties
of the {vertically-emergent ($\mu=1.00$)} synthetic profile  features are related to the physical properties of the
{individual columns of the 3D} model atmosphere at the corresponding heights.
We present results for the K line only as they are similar for the H line.

We use the following notations.
The speed of light is $c$.
The vertical velocity is $\varv_\text{Z}(z)$ and it depends on height $z$.
The central wavelength of the K line is $\lambda_0$.
For K$_\text{2V}$, K$_3$, and K$_\text{2R}$, we denote their wavelengths
$\lambda(\text{K}_\text{2V})$, $\lambda(\text{K}_3)$, and
$\lambda(\text{K}_\text{2R})$.
The same notation in parentheses is used for the emergent intensity $I$ and the
corresponding brightness temperature $T_\text{b}$.
The formation height of K$_3$ is $z_3 \equiv z(\text{K}_3)$.
Similarly, the averaged formation height of K$_2$ is
\begin{equation} \label{eq:z_2}
  z_2
    =
    \tfrac{1}{2}
    \bigl[
      z_\text{2V} + z_\text{2R}
    \bigr]
    \equiv
    \tfrac{1}{2}
    \bigl[
      z(\text{K}_\text{2V}) + z(\text{K}_\text{2R})
    \bigr].
\end{equation}
The Doppler shift of K$_3$ is
\begin{equation} \label{eq:v_3}
  \varv_3
    =
    c \dfrac{ \Delta\lambda_3 }{ \lambda_0 }
    \equiv
    c \dfrac{ \lambda_0 - \lambda(\text{K}_3) }{ \lambda_0 }.
\end{equation}
The averaged Doppler shift of K$_2$ is
\begin{equation} \label{eq:v_2}
  \varv_2
    =
    c
    \dfrac{
      \lambda_0 - \tfrac{1}{2}
      \bigl[ \lambda(\text{K}_\text{2V}) + \lambda(\text{K}_\text{2R}) \bigr]
    }{
      \lambda_0
    }.
\end{equation}
The peak-to-peak distance or the peak separation is
\begin{equation} \label{eq:Dv_2}
  \Delta\varv_2
    =
    c \dfrac{ \Delta\lambda_2 }{ \lambda_0 }
    \equiv
    c \dfrac{
      \lambda(\text{K}_\text{2R}) - \lambda(\text{K}_\text{2V})
    }{ \lambda_0 }.
\end{equation}
The averaged vertical velocity at peaks is
\begin{equation} \label{eq:v_Z2}
  \langle \varv_\text{Z} \rangle_2
    =
    \tfrac{1}{2}
    \bigl[
      \varv_\text{Z}( z_\text{2V} ) + \varv_\text{Z}( z_\text{2R} )
    \bigr].
\end{equation}
The maximum amplitude of the vertical velocity
\begin{equation} \label{eq:Dv_Z}
  \Delta \varv_\text{Z}
    =
    \max_{z_2 \leq z \leq z_3} v_\text{Z}(z) -
    \min_{z_2 \leq z \leq z_3} v_\text{Z}(z)
\end{equation}
is measured between $z_2$ and $z_3$, that is, the range of heights where the
central part of the profile between the emission peaks is formed.
In the same range of heights we define the mean vertical velocity
\begin{equation} \label{eq:mean_v_Z}
  \newcommand*{\dd}{ \ensuremath{ \mathrm{d} } } 
  \langle \varv_\text{Z} \rangle_{2{-}3}
    =
    \dfrac{ 1 }{ z_3 - z_2 }
    \int_{z_2}^{z_3} \!\varv_\text{Z}(\zeta) \,\dd\zeta.
\end{equation}
The peak asymmetry $A$ is the same as in Eq.\ \eqref{eq:peak-asymmetry}.

\subsubsection{Velocities}
\label{sec:feature-velocities}

\begin{figure*}
  \sidecaption
  \begin{minipage}[b]{120mm}
    \includegraphics[%
      width = 60mm,%
      trim  = 0pt 1pt 0pt 1pt,%
      clip  = true%
    ]{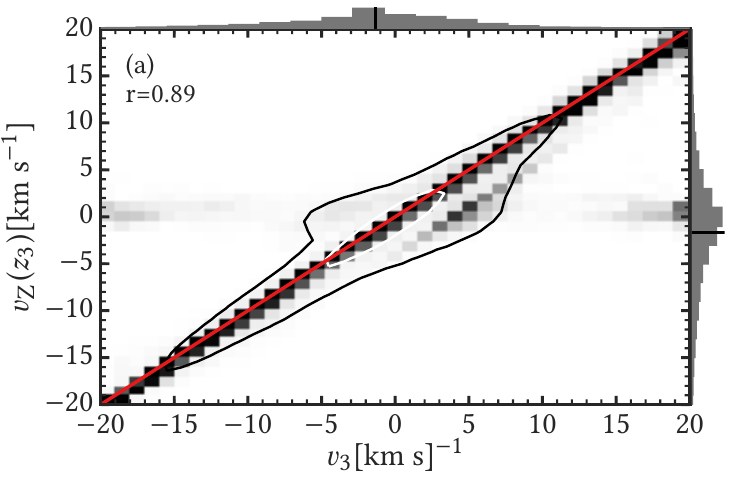}\hfil
    \includegraphics[%
      width = 60mm,%
      trim  = 0pt 2pt 0pt 2pt,%
      clip  = true%
    ]{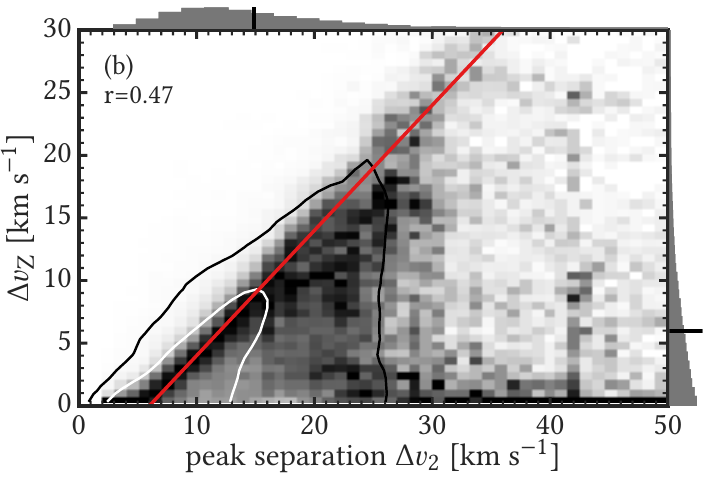}
    \\
    \includegraphics[%
      width = 60mm,%
      trim  = 0pt 1pt 0pt 1pt,%
      clip  = true%
    ]{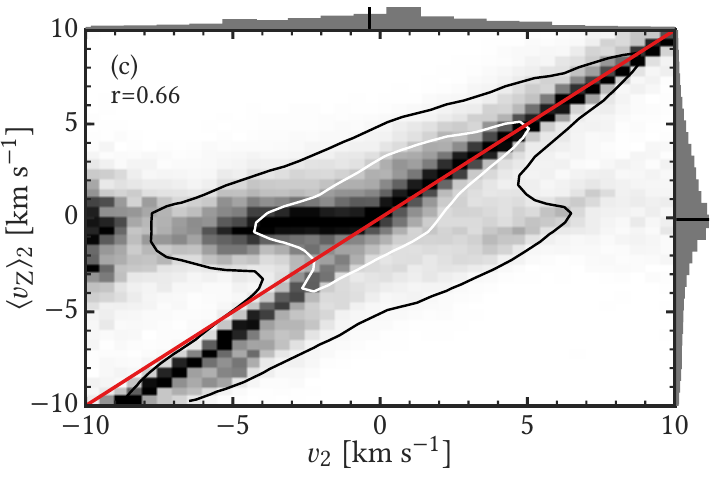}\hfil
    \includegraphics[%
      width = 60mm,%
      trim  = 0pt 1pt 0pt 1pt,%
      clip  = true%
    ]{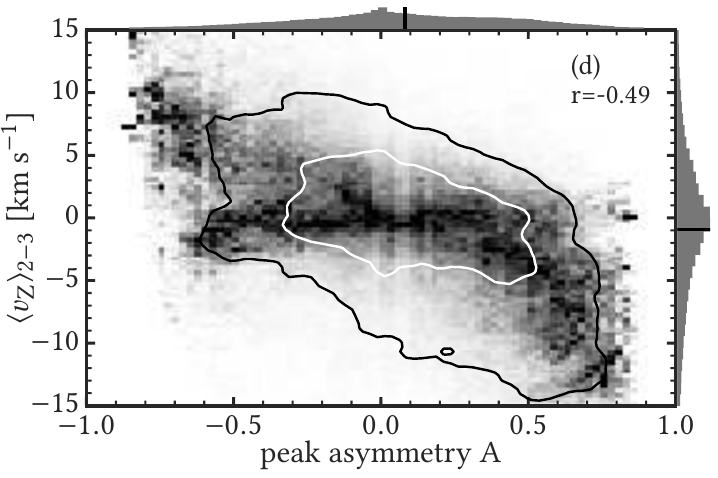}%
  \end{minipage}
  \caption{%
    Correlations of observable K line profile properties for diagnosing
    vertical velocities in the chromosphere from Model 2.
    Panel~a): the K$_3$ Doppler shift against the vertical velocity at the K$_3$
    formation height.
    Panel~b): the K$_2$ peak separation against the maximum vertical velocity
    amplitude between K$_2$ and K$_3$ formation heights.
    Panel~c): the average K$_2$ Doppler shift against the averaged vertical
    velocity at the K$_2$ formation heights.
    Panel~d): the emission peak asymmetry against the mean vertical velocity
    between the K$_2$ and K$_3$ formation heights.
    The panels follow the same format as Fig.~\ref{fig:fig1_1dprd_vs_3dprd}.%
  }
  \label{fig:fig_zmin}
\end{figure*}
\citetads{2013ApJ...772...90L} 
showed that the \ion{Mg}{II} h and k lines are good for tracing the
line-of-sight velocities in the chromosphere through the h$_3$/k$_3$ or the
h$_2$/k$_2$ features.
We test whether the same is true for the \ion{Ca}{II} H and K lines.

We examined whether the K$_3$ Doppler shift $\varv_3$ (Eq.\ \ref{eq:v_3})
corresponds to the vertical velocity $v_\text{Z}(z_3)$ at the K$_3$ formation
height $z_3$.
Panel~a) in Fig.\ \ref{fig:fig_zmin} shows this is true with a very strong
correlation.
Two spurious spots outside of the main distribution resulted from K$_3$ misfits
in complex profiles with more than two emission peaks.
The wavelength position of K$_3$ is a very accurate probe for velocities
in the upper chromosphere.

We studied how the K$_2$ peak separation $\Delta \varv_2$ (Eq.\ \ref{eq:Dv_2})
is related to the corresponding maximum amplitude of the vertical velocity
$\Delta\varv_\text{Z}$ (Eq.\ \ref{eq:Dv_Z}).
Panel~b) in Fig.\ \ref{fig:fig_zmin} shows a decent correlation for
$\Delta \varv_2 < 20$~km\,s$^{-1}$. {An example is given in panel (a) of Figure \ref{fig:four_panel_1}, where a 11 km\,s${^{-1}}$ K$_2$ peak separation corresponds to $\Delta\varv_\text{Z}$=5 km\,s${^{-1}}$.}
A larger separation of the K$_2$ peaks is caused by the deep chromospheric heating
discussed in Section~\ref{sec:deep_chromo} and is not dependent on the velocity amplitudes.

We related the averaged K$_2$ Doppler shift $\varv_2$ (Eq.\ \ref{eq:v_2}) with
the averaged vertical velocity $\langle\varv_\text{Z}\rangle_2$ at K$_2$
(Eq.\ \ref{eq:v_Z2}).
Panel~c) in Fig.\ \ref{fig:fig_zmin} shows that this is a good velocity
diagnostic for the middle chromosphere, especially for strong velocities.
The distribution shows a number of points sticking out towards the left at $\langle\varv_\text{Z}\rangle_2=0$. They are mainly caused by misidentifications of one or both of the \Ktwo\ peaks. We note that in the simulations one has access to the formation heights of the peaks, while this is not the case for observations. Observationally, $\varv_2$ can thus be used to estimate the vertical velocity in the chromosphere at the peak {formation} heights,  but it is not possible to estimate the formation heights themselves.

Finally, we related the peak asymmetry $A$ (Eq.\ \ref{eq:peak-asymmetry}) with
the mean vertical velocity $\langle\varv_\text{Z}\rangle_{2{-}3}$
(Eq.\ \ref{eq:mean_v_Z}).
In the quiet Sun observations, the peak asymmetry is mostly positive
indicating in the chromosphere a particular type of downflows that follow upward
passages of shock waves
\citepads{1997ApJ...481..500C}. 
Panel~d) in Fig.\ \ref{fig:fig_zmin} shows an anti-correlation of $A$ with
$\langle\varv_\text{Z}\rangle_{2{-}3}$, which means that the blue emission peak
becomes stronger than the red emission peak if material in the middle-upper
chromosphere is mainly moving down and vice versa.
This dependence is almost linear for small velocities but then saturates
for large ones.

\subsubsection{K$_2$ intensities}
\label{sec:K2-temperatures}

\begin{figure*}
  \begin{minipage}{\textwidth}
    \includegraphics[%
      width = \textwidth,%
      trim  = 2pt 4pt 3pt 2pt,%
      clip  = true%
    ]{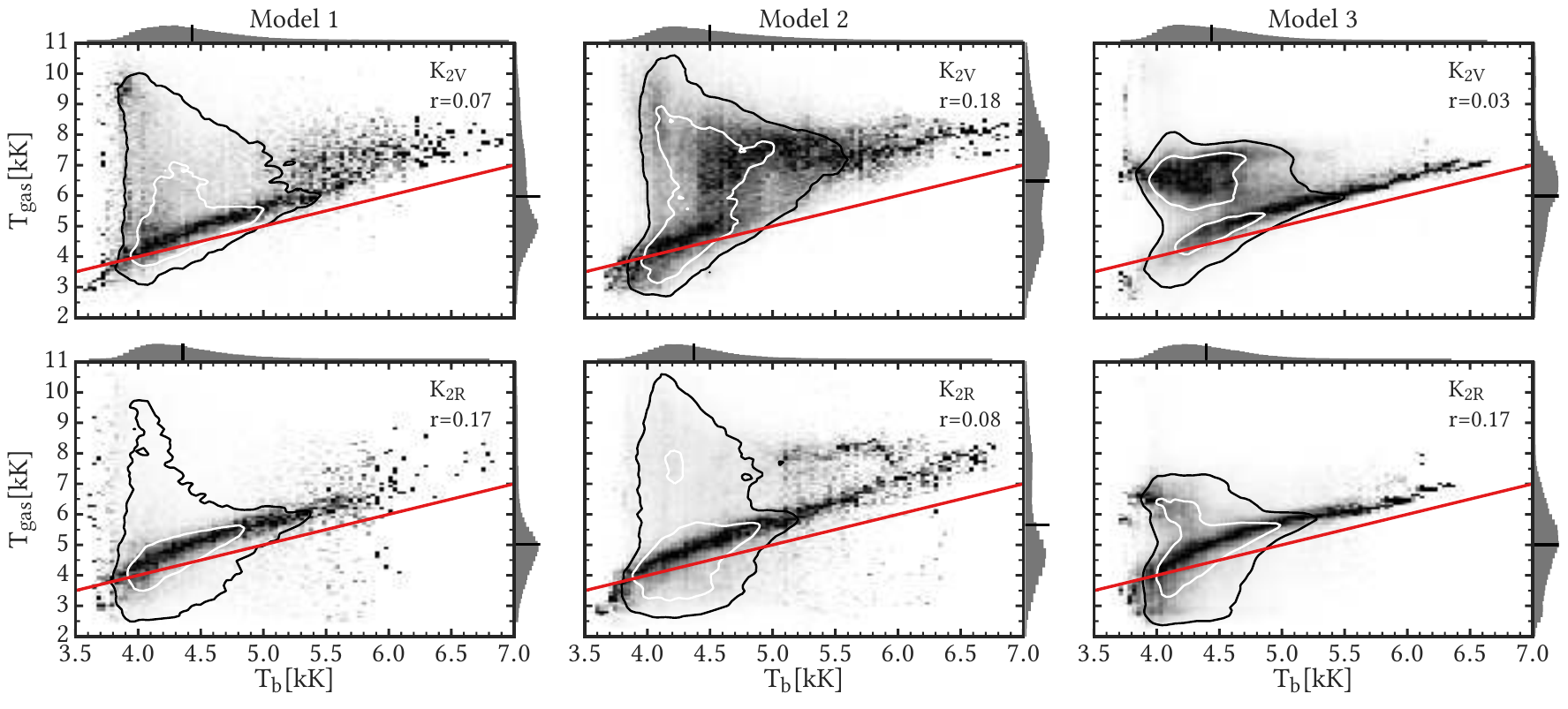}%
  \end{minipage}%
  \caption{%
    Calibrations of the observed brightness temperature at K$_\text{2V}$
    (\emph{top row}) and K$_\text{2R}$ (\emph{bottom row}) for measuring the
    gas temperature at the related formation heights.
    Results are shown for Model~1, 2, and 3
    (\emph{left, center, and right column}).
    The panels follow the same format as Fig.~\ref{fig:fig1_1dprd_vs_3dprd}.%
  }
  \label{fig:fig7_temp}
\end{figure*}
The emission peaks of the h and k lines of \ion{Mg}{II}
\citepads{2013ApJ...772...90L} 
demonstrate a correlation of their brightness temperature $T_\text{b}$ with the
gas temperature $T_\text{gas}$ at the corresponding formation heights.
We investigated the validity of this relation for the emission peaks of the \CaIIHK\ lines.

Figure~\ref{fig:fig7_temp} shows correlations between the brightness temperature
$T_\text{b}(\text{K}_\text{2V})$ at K$_\text{2V}$ or
$T_\text{b}(\text{K}_\text{2R})$ at K$_\text{2R}$ and the related gas
temperature $T_\text{gas}(z_\text{2V})$ or $T_\text{gas}(z_\text{2R})$.
For each model atmosphere there is a certain range of $T_\text{b}$ where this
relation is valid, therefore the peaks can probe the gas temperature in the
chromosphere.

In Model~1, this range is 4.5--6~kK.
Below 4.5~kK, the peak intensity gets set by the scattered radiation in the
middle chromosphere and the distribution spreads out decreasing the correlation.
Intensities at both peaks underestimate the gas temperature by 0.5--1~kK.

In Model~2, the range of the linear correlation is 4.7--6.8~kK.
The blue peak intensity underestimates the gas temperature by $\sim 2$~kK, while the
red peak intensity underestimates it by 1~kK.
{The difference between red and blue peak} is caused by much stronger velocity fields in this model, {which causes large variations of the opacity along the line of sight (For examples see Figure \ref{fig:four_panel_1})}.

In Model~3, the validity range is 4.7--6.4~kK.
This model shows the most accurate linear dependence with the smallest spread.
There are side secondary clusters of points at 3.7--4.6~kK for K$_\text{2V}$ and at 3.7--4.2~kK for
K$_\text{2R}$ where there is no correlation and intensity is controlled by the
scattered radiation.

In all models, the K$_\text{2R}$ peak shows a stronger correlation
than the K$_\text{2V}$ peak.


We test the correlations by computing them using the line profiles
with a single emission peak only, which constitute only 16\% of the entire
population.
We obtain very similar distributions, therefore we conclude that these types
of correlations are not sensitive to our algorithm for the profile feature
classification.

We test this relation for the K$_3$ feature as well.
The line core is strongly scattering and its brightness temperature is not correlated to the local gas temperature.
Therefore, the \Kthree\ brightness temperature cannot be used to measure the gas temperature in the
chromosphere.

\subsubsection{K$_1$ intensities}
\label{sec:K1-temperatures}

\begin{figure}
  \begin{minipage}{\columnwidth}
    \includegraphics[%
      width = \columnwidth,%
      trim  = 0pt 0pt 0pt 0pt,%
      clip  = true%
    ]{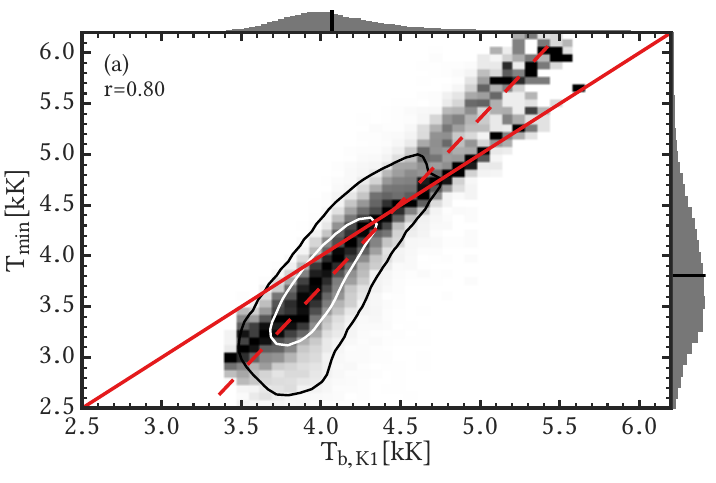}
    \\
    \includegraphics[%
      width = \columnwidth,%
      trim  = 0.2pt 0pt 0pt 2pt,%
      clip  = true%
    ]{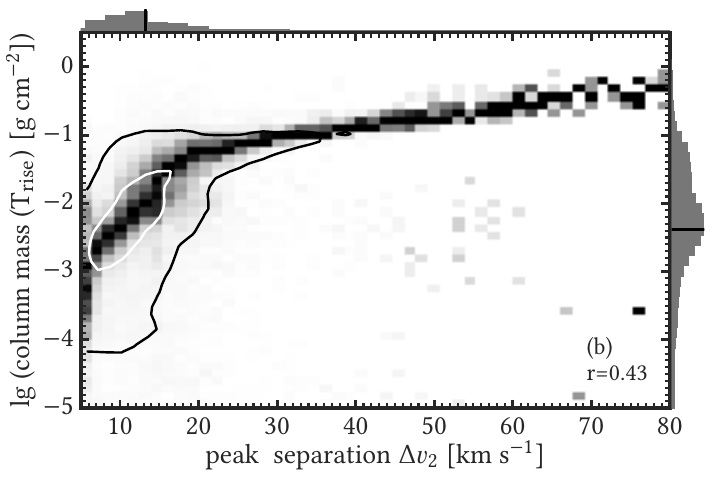}%
  \end{minipage}%
  \caption{%
    Observed properties of the lower chromosphere from Model 1.
    Panel~a): the averaged brightness temperature at K$_1$ against the gas
    temperature at the temperature minimum.
    {The red solid line is $y = x$ and the} dashed line is the 3rd-degree polynomial fit.
    Panel~b): the K$_2$ peak separation against the column mass above the
    temperature rise.
    The panels follow the same format as Fig.~\ref{fig:fig1_1dprd_vs_3dprd}.%
  }
  \label{fig:fig18_heat}
\end{figure}
The K$_1$ minima are formed between the upper photosphere and the lower
chromosphere.
There, they are caused by either a global temperature minimum or one of the local minima.
The line source function at such heights is often still partially coupled to the Planck
function.
Therefore the K$_1$ minima can be used to diagnose the temperature in the temperature minimum.

\citetads{1975ApJ...199..724S} 
investigated  how the K$_1$ intensity relates
to the global temperature minimum in the 1D HRSA model atmosphere. They found that the brightness temperature at K$_1$ was lower
than the gas temperature at the minimum.

We apply their approach to measure the temperature minimum in our 3D Bifrost model
atmospheres.
Such 3D R-MHD model atmospheres have a very complicated temperature structure
that is not as easy to classify as it can be done with traditional 1D
hydrostatic model atmospheres that typically have a single well-defined temperature minimum around a height of 500 km.

Since the line source function is frequency-dependent in the line wings due to
PRD, the K$_\text{1V}$ or K$_\text{1R}$ features can be formed at different
formation heights. Therefore, we first determine the formation height of each feature and then search in a height interval of 200 km around these formation heights for the deepest temperature minimum. We define the global temperature minimum as the height with the lowest gas temperature
{obtained} either from K$_\text{1V}$ or K$_\text{1R}$.

{Panel a) in Fig.}~\ref{fig:fig18_heat} shows how the averaged brightness temperature
$T_\text{b}(\text{K}_1)$ at K$_1$ relates to the gas temperature at the minimum.
The K$_1$ brightness temperature overestimates the minimum's temperature at low \Kone\ intensities.
From $T_\text{b}(\text{K}_1) = 4.4$~kK and above there are two arms in the
distribution.
One, along the red solid line, shows a tight linear correlation meaning that the
radiation temperature and the gas temperature are well coupled.
The other, along the red dashed line, underestimates the minimum's temperature. The atmospheric columns that produce the line   along the red {solid} line all have very wide \Kone\ separations and formation heights so low in the atmosphere that the source function is still strongly coupled to the Planck function.

In all three models, this global correlation is very strong.
We conclude that K$_1$ could be used to assess the temperature of the temperature minimum using {panel a) in }Fig.~\ref{fig:fig18_heat} as a calibration. The bifurcation of the distribution above $T_\text{b}(\text{K}_1) = 4.6$~kK adds some ambiguity, however.

\subsubsection{\Ktwo\ peak separation}
\label{sec:deep_chromo}

As we noted for the distribution on panel~b) in Fig.\ \ref{fig:fig_zmin}, there
is no correlation between the K$_2$ peak separation and the maximum velocity
amplitude in the middle-upper chromosphere if the former is more than
20~km\,s$^{-1}$.
By applying the Eddington-Barbier approximation, we conjecture that a separation between the K$_2$ peaks significantly larger than  20~km\,s$^{-1}$ indicates that there is a temperature increase already much deeper in the atmosphere than typical for our simulations {(an example of a profile is given in panel (d) of Figure \ref{fig:four_panel_1})}.


We identify the location of this temperature increase assuming that at the formation height $z_2$ of K$_2$
the source function is decreasing when moving deeper into the atmosphere.
If this is true, we stop at height where the source function is 90\% of its
value at $z_2$.
From this height up we integrate the density to obtain the column mass where the
temperature rise occurs.
If K$_2$ is misidentified or if the source function is only increasing with increasing depth, we
discard this profile from the sample.

{Panel b) in Fig.}~\ref{fig:fig18_heat} shows that K$_2$ peak separations above 25 km~s$^{-1}$ are indeed associated with a deeply located temperature increase (i.e., at high column mass). The correlation is very tight.  We note that in our model atmospheres we only have a very small fraction of profiles that
show this effect. Nevertheless, because the effect is based on simple radiation transfer properties we expect that very wide K$_2$ peak separations can be evidence of a deep chromospheric temperature rise also in observations. However, because the observed and simulated line profiles show substantial differences, we also note that observed wide peak separations can be caused by effects not present in our models.

\section{Summary and conclusions}
\label{sec:discussion}

We investigated the formation and the diagnostic value of the \ion{Ca}{II} H
and K lines through observations and numerical modeling.

We modeled the \ion{Ca}{II} spectrum by considering the non-LTE, the 3D~RT, and
{PRD}/XRD effects together. 3D~RT effects are important in the cores of the lines,
while the PRD/XRD effects mostly influence the wings.
A joint treatment of all three effects is important to obtain correct synthetic
intensities.

We computed synthetic line profiles in 3D non-LTE including XRD from snapshots from three different radiation-MHD models computed with the Bifrost code.



We compared the synthetic spatially-averaged spectrum with a standard solar
atlas and our SST/CHROMIS observations.
None of the model atmospheres reproduces the observed spectral profiles.
In Model~2, the emission peaks are usually blended at disk-center
and become separated only towards the limb.
Model~1 and Model~3 reproduce the two emission peaks.
All three models have a too low wavelength separation between the K$_1$ and K$_2$ features.
We conclude that something is missing in all models that is
responsible for the broader observed profiles. Two possibilities for this are a lack of motions at scales {smaller than the photon mean free path (''microturbulence'', 1D modeling shows that amplitudes of ~5 km s$^{-1}$ are sufficient)}, and too weak heating processes in the lower chromosphere (see {Fig.~\ref{fig:eb_four_panel}, {panel (d) of Fig.~\ref{fig:four_panel_1}}, and} the lower panel of Fig.~\ref{fig:fig18_heat}). 

The same behaviour has been reported for the h and k lines of \ion{Mg}{II}
\citepads{2013ApJ...772...89L, 
          2013ApJ...772...90L,
          2013ApJ...778..143P} 
and \ion{C}{II} lines
\citepads{2015ApJ...811...80R} 
in Model~1.
The models also predict a too low \Ktwo\ radiation temperature.

We compared the center-to-limb variation with Model~1, Model~2, and observations
performed at Sacramento Peak
\citepads{1968SoPh....3..164Z}. 
The models reproduce the observed trends in variation of the intensity of the line features and
the K$_1$\ and K$_2$ separations, but do not reproduce their absolute values.

We investigated several diagnostic possibilities of the H and K lines.
The H$_3$/K$_3$  features trace the vertical
velocity in the upper chromosphere.
Furthermore, the \Ktwo\ peak separation correlates
with the velocity difference between the K$_2$ and K$_3$ formation heights for peak separations below 20~km\,s$^{-1}$.
For larger peak separation we find a good correlation with the column mass where the chromospheric temperature rise occurs.
The \Ktwo\ asymmetry can be used to measure the average velocity between the line core
and the emission peaks.

The models predict a too low peak separation, and the correlation that we found thus point towards, on average, too low vertical velocities fields and a too high location of the chromospheric temperature rise in the models.

The brightness temperature of H$_2$/K$_2$ and H$_1$/K$_1$ can probe the
local conditions in the upper photosphere to the middle chromosphere.
We showed that the brightness temperature of H$_1$/K$_1$ correlates with the temperature in a local or global
temperature minimum along the line of sight.
The brightness temperature of the H$_2$/K$_2$ features correlates with the gas temperature at their formation heights, especially for high temperatures. The offset between the gas temperature and the K$_2$ brightness temperature is somewhat different in all three models, which means that the temperature estimates derived from the brightness temperature have an uncertainty {in the range of $0.5-2$kK.}

The H and K lines of \ion{Ca}{II}  have similar formation
properties similar as the h and k lines of \ion{Mg}{II}.
The main difference is the larger formation heights of the h and k lines. Within magnetic elements this height difference is small, but in the simulated internetwork regions the difference can be up to 2\,000~km.

We studied three different model atmospheres that span a large variation of
physical conditions, and all models produce similar correlations between
observables and the atmospheric parameters.
We therefore believe that the correlations presented in this paper will
be valid in the quiet Sun.
However, these correlations might not be valid under different circumstances,
for example in active regions.



{The simulations that we use here do not include the Hall term and ambipolar diffusion resulting from the interaction of ions and neutral particles.}
\citetads{2012ApJ...753..161M} 
has shown that inclusion of those effects in 2.5D models leads to enhanced dissipation of
magnetic free energy, which, { in turn,} leads to an increase in heating in the
chromosphere.

In addition,
\citetads{Martinez-Sykora1269}
showed that ion-neutral effects in a different 2.5D simulation with a larger spatial extent and higher spatial resolution, produces structures that have the same properties as type II spicules
\citepads{2007PASJ...59S.655D}.
These spicules are notably absent from our simulations, and might play a role in setting the average properties of the \CaIIHK\ line profiles.

All three snapshots that we use are computed with a different equation of state. The equation of state has a large impact on the density and temperature
structure in the chromosphere and transition region
\citepads{2007A&A...473..625L,2016ApJ...817..125G}.
We can therefore not draw any conclusions with respect to the effect of EOS or spatial resolution on the line profiles. 

Radiative transfer computations as described in this manuscript should be performed on new 3D simulations with a higher resolution, an equation of state including non-equilibrium ionization of both hydrogen and helium, and including the effects of ion-neutral interactions.

\begin{acknowledgements}
  We thank David B\"uhler and Jayant Joshi for acquiring the quiet-Sun data.
  The Swedish 1-m Solar Telescope is operated on the island of La Palma by the
  Institute for Solar Physics of Stockholm University in the Spanish
  Observatorio del Roque de los Muchachos of the Instituto de Astrof\'isica de
  Canarias.
  The computations were performed on resources provided by the Swedish National
  Infrastructure for Computing (SNIC) at the High Performance Computing Center
  North at Ume\aa\ University and the PDC Centre for High Performance Computing
  (PDC-HPC) at the Royal Institute of Technology in Stockholm.
  JdlCR is supported by grants from the Swedish Research Council (2015-03994),
  the Swedish National Space Board (128/15) and the Swedish Civil Contingencies Agency (MSB).
  This research was supported by the CHROMOBS and CHROMATIC grants of the Knut
  och Alice Wallenberg foundation and by the Research Council of Norway through the grant
  "Solar Atmospheric Modelling" and through grants of computing time from the Programme for Supercomputing.

\end{acknowledgements}

\bibliographystyle{aa} 
\bibliography{article}
\end{document}